\documentclass[
  subscriptcorrection,
  upint,
  varvw,
  barcolor=Goldenrod3,
  balance,
  hyphenate,
  french
]{asmejour}

\usepackage[utf8]{inputenc}
\usepackage{textcomp,algorithm}
\usepackage[italicComments=true,noEnd=false,indLines=true]{algpseudocodex}

\newcommand{\RR}{\mathbb{R}}

\newcommand{\D}{\mathcal{D}}
\renewcommand{\H}{\mathcal{H}}
\newcommand{\K}{\mathcal{K}}

\newcommand{\M}{\mathcal{M}}
\newcommand{\G}{\mathcal{G}}
\newcommand{\U}{\mathcal{U}}

\hypersetup{%
	pdfauthor={Kartik Loya},                       		   	
	pdftitle={Koopman Operator Framework for Modeling and Control of Off-Road Vehicle on Deformable Terrai},                  	
	pdfkeywords={Koopman operator, Subspace Identification, Offroad vehicle dynamics, deformable terrain, Bekker-Wong-Reece model},
	pdfsubject = {Koopman operators for off-road ground robots},			
    }

\JourName{Autonomous Vehicles and Systems}

\begin{document}


\SetAuthorBlock{Kartik Loya}{Department of Mechanical Engineering,\\
Clemson University, \\
Clemson, SC 29607 USA \\
   email: kloya@clemson.edu} 


\SetAuthorBlock{Phanindra Tallapragada\CorrespondingAuthor{}}{%
Department of Mechanical Engineering,\\
Clemson University, \\
Clemson, SC 29607 USA \\
email: ptallap@clemson.edu
}

\title{Koopman Operator Framework for Modeling and Control of Off-Road Vehicle on Deformable Terrain}


\begin{abstract}This work presents a hybrid physics-informed and data-driven modeling framework for predictive control of autonomous off-road vehicles operating on deformable terrain. Traditional high-fidelity terramechanics models are often too computationally demanding to be directly used in control design.  Modern Koopman operator methods can be used to represent the complex terramechanics and vehicle dynamics in a linear form. We develop a framework whereby a Koopman linear system can be constructed using data from simulations of a vehicle moving on deformable terrain. For vehicle simulations, the deformable-terrain terramechanics are modeled using Bekker–Wong theory, and the vehicle is represented as a simplified five-degree-of-freedom (5-DOF) system.  The Koopman operators are identified from large simulation datasets for sandy loam and clay using a recursive subspace identification method, where Grassmannian distance is used to prioritize informative data segments during training. The advantage of this approach is that the Koopman operator learned from simulations can be updated with data from the physical system in a seamless manner, making this a hybrid physics-informed and data-driven approach. Prediction results demonstrate stable short-horizon accuracy and robustness under mild terrain-height variations. When embedded in a constrained MPC, the learned predictor enables stable closed-loop tracking of aggressive maneuvers while satisfying steering and torque limits.  

\end{abstract}

\date{Version \versionno, \today. \\ DISTRIBUTION STATEMENT A. Approved for public release; distribution is unlimited. OPSEC \# 10375. }

\maketitle 


\section{Introduction}
Off-road autonomy for ground vehicles is fundamentally more challenging than on-road autonomy because the dominant uncertainty stems from physics of contact with the terrain, rather than just perception or mapping. Friction-based tire models are often adequate on rigid surfaces with stable contact conditions, deformable off-road terrain can break this assumption because deformation couples traction and steering to terrain strength and slip. When a tire runs on deformable terrain, the ground can compact and shear, the wheel can sink, and the effective contact geometry changes continuously. As a result, the longitudinal and lateral forces that govern acceleration and steering depend strongly on soil properties, normal load, slip, and vehicle state in ways that are difficult to predict using standard vehicle models used for on-road autonomy. High fidelity methods such as discrete element simulations and continuum granular models can capture soil flow and failure with high realism, but their computational cost is very high for control design \cite{huCalibrationExpeditiousTerramechanics2024}. Classical terramechanics models such as Bekker and Wong-Reece are more structured and use pressure sinkage and shear stress relations to capture many of these effects and form the foundation for predicting mobility on soft soil; even so their evaluation can be expensive because they often require contact patch discretization, stress computation, force integration, and iterative updates to satisfy normal load balance and slip-dependent shear behavior, which makes repeated calls inside online planning, estimation, and MPC difficult at real-time rates \cite{heReviewTerramechanicsModels2019}.

Hence, the challenge goes beyond modeling accuracy of terramechanics since many deformable-terrain models are expensive to evaluate in real time.  At the same time, rigid-surface approximations are often insufficient on soft soil as force prediction errors can degrade tracking and may lead to slip, sinkage, and reduced controllability during high-speed or aggressive maneuvers.  This has motivated control-oriented terramechanics approximations and learned mobility models that trade fidelity for speed \cite{barthelmesTerRATerramechanicsRealtime2018}. Recent terrain-adaptive N-MPC formulations incorporate deformable terrain effects through neural networks to improve short-horizon accuracy, see for example \cite{dallasTerrainAdaptiveTrajectory2021}. But computationally efficient control models remain a significant challenge for off-road autonomy.

In this paper, we identify a computationally efficient and updateable Koopman-based linear predictor for off-road vehicle dynamics on deformable terrain. We use simulation to efficiently and safely generate large and diverse datasets of vehicle motion on soft soil. The datasets are generated using a simplified 5-dof vehicle model coupled with Bekker–Wong–Reece terramechanics model. From these datasets, we then identify soil-specific lifted linear Koopman models that retain the dominant nonlinear wheel–terrain effects while enabling short-horizon state prediction in real-time, suitable for receding-horizon control and trajectory planning. Although Bekker–Wong–Reece simulations are used here for demonstration, the overall learning and control pipeline is agnostic to the particular simulator, terramechanics formulation, or vehicle platform. To make the Koopman predictor updateable and scalable, we use a recursive subspace identification approach with Gaussian processes as basis functions that enable incremental learning from data segments, following our previous work \cite{LBT_dsc_2023, LT_mecc_2024} and the procedure summarized in Section~\ref{sec:3}. Our method avoids building and factorizing large Hankel and regression matrices as the dataset grows, as the algorithm maintains a compact subspace representation that can be updated efficiently with the addition of new segments. The update can be applied in batches when additional maneuvers or operating regimes are introduced, and a Grassmannian distance test is used to accept only information-rich segments while rejecting repeated or redundant data. In this way, physics-informed Koopman operators learned from simulation can be refined using large datasets without incurring the memory and computational cost of full re-identification.

The remainder of the paper is organized as follows. Section \ref{sec:2} presents the coupled vehicle–terrain simulation model, including the 5-DOF vehicle dynamics and the Bekker–Wong–Reece wheel–soil interaction used to generate training data. Section \ref{sec:3} describes the Koopman identification pipeline and the recursive subspace identification procedure used to learn lifted linear predictors from large datasets. Section \ref{sec:4} reports prediction results on sandy loam and soils, including short-horizon accuracy and robustness tests under mild terrain-height variations. In Section \ref{sec:kmpc}, we use the learned predictor in a constrained Koopman-MPC and demonstrate closed-loop tracking of aggressive maneuvers under steering and torque limits. Section \ref{sec:6} highlights the need for terrain-specific Koopman operators and compares performance across soil types, and Section \ref{sec:7} concludes with key findings and future directions.





\section{Dynamic Modeling of the Vehicle–Terrain System}\label{sec:2}
The vehicle model considered in this work combines a bicycle formulation for steering and longitudinal dynamics with a half-car suspension model to capture the vertical and pitch dynamics of the vehicle. The interaction between these subsystems arises through the wheel–terrain contact forces, which depend on the time-varying normal loads influenced by suspension motion and terrain deformability. To accurately represent these effects on soft ground, the Bekker–based terramechanics model is integrated to predict normal and shear stresses at the wheel–terrain interface. The following sections present the governing equations for each subsystem and illustrate how their coupling yields a comprehensive, yet computationally efficient model for off-road vehicle simulation  \cite{rajamaniVehicleDynamicsControl2006,taheriTechnicalSurveyTerramechanics2015,heReviewTerramechanicsModels2019}.

\subsection{Single-track (bicycle) dynamics model}

\begin{figure}[h]
        \centering
        \includegraphics[width=0.4\textwidth, keepaspectratio]{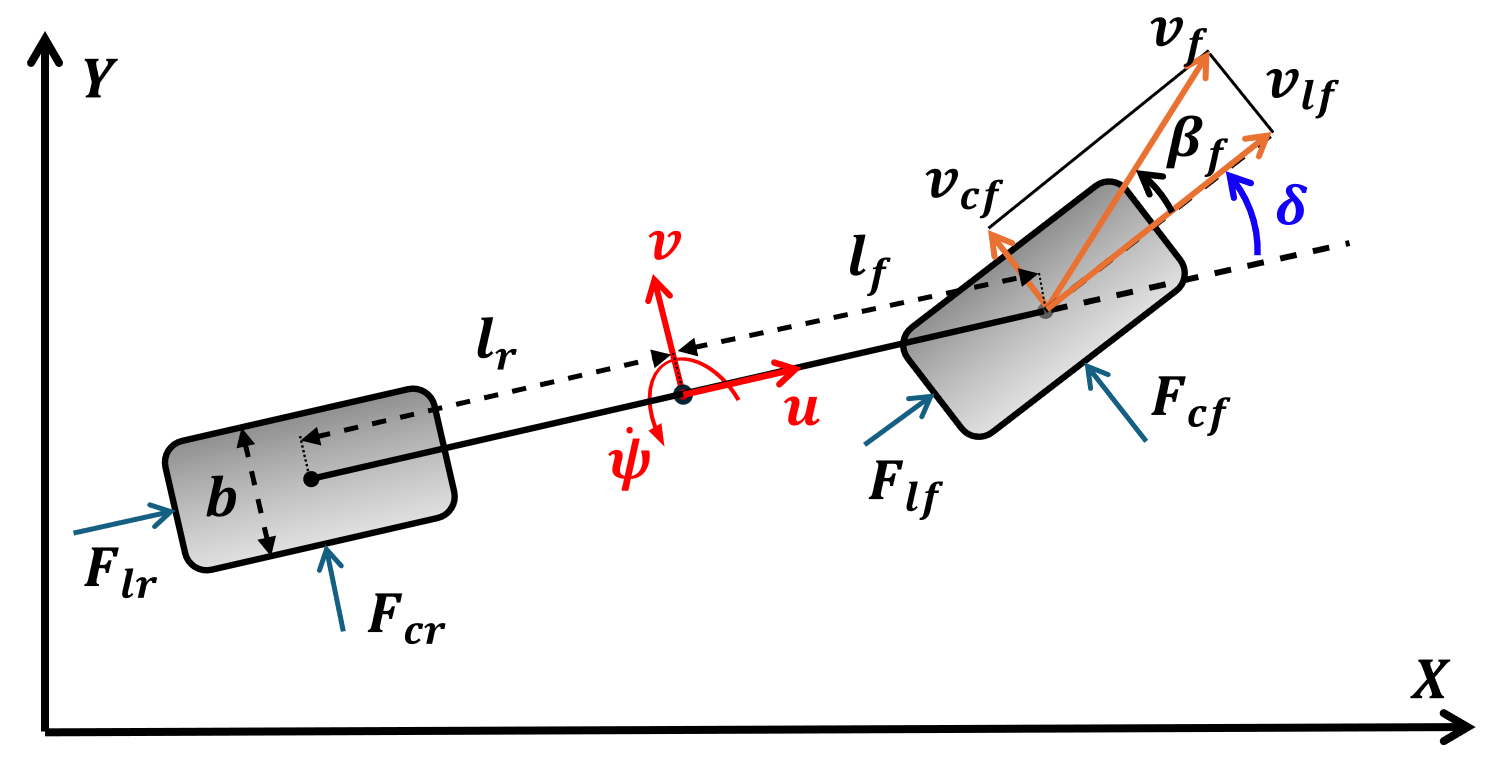}
        \caption{Longitudinal dynamics using single-track (bicycle) dynamic model}\label{Fig:longitudinal}
\end{figure}
The planar motion of the vehicle is represented using a single-track dynamic (bicycle) model \cite{falconePredictiveActiveSteering2007}, which captures the lateral, longitudinal, and yaw dynamics through a simplified two-wheel abstraction. The single-track assumption combines the left and right wheels of each axle into a single equivalent wheel, allowing compact representation of tire forces while preserving key characteristics. The equations of motion are written by summing forces and moments in a body-fixed frame of reference,
\begin{subequations}\label{Eq:longitudinal}
    \begin{align}\label{Eq:longitudinal_x}
     m\dot{u} =& ~m(v \dot{\psi} \cos{\theta} + \dot{z} \dot{\theta}) + F_{lf}\cos{\delta} - F_{cf}\sin{\delta} ~+ \\ & F_{lr} -f_{ax}, \nonumber \\ \label{Eq:longitudinal_y}
    m\dot{v} =& ~m(-u \dot{\psi} \cos{\theta} + \dot{z} \dot{\theta} \sin{\theta}) + F_{lf}\sin{\delta} - F_{cf}\cos{\delta} ~+\\ & F_{cr} - f_{ay}, \nonumber \\  \label{Eq:longitudinal_psi}
    I_z\ddot{\psi} =& ~l_f(F_{lf}\sin{\delta} + F_{cf} \cos{\delta}) - F_{cr} l_r. 
\end{align}
\end{subequations}

\begin{figure}[t]
\centering\includegraphics[width=0.45\textwidth]{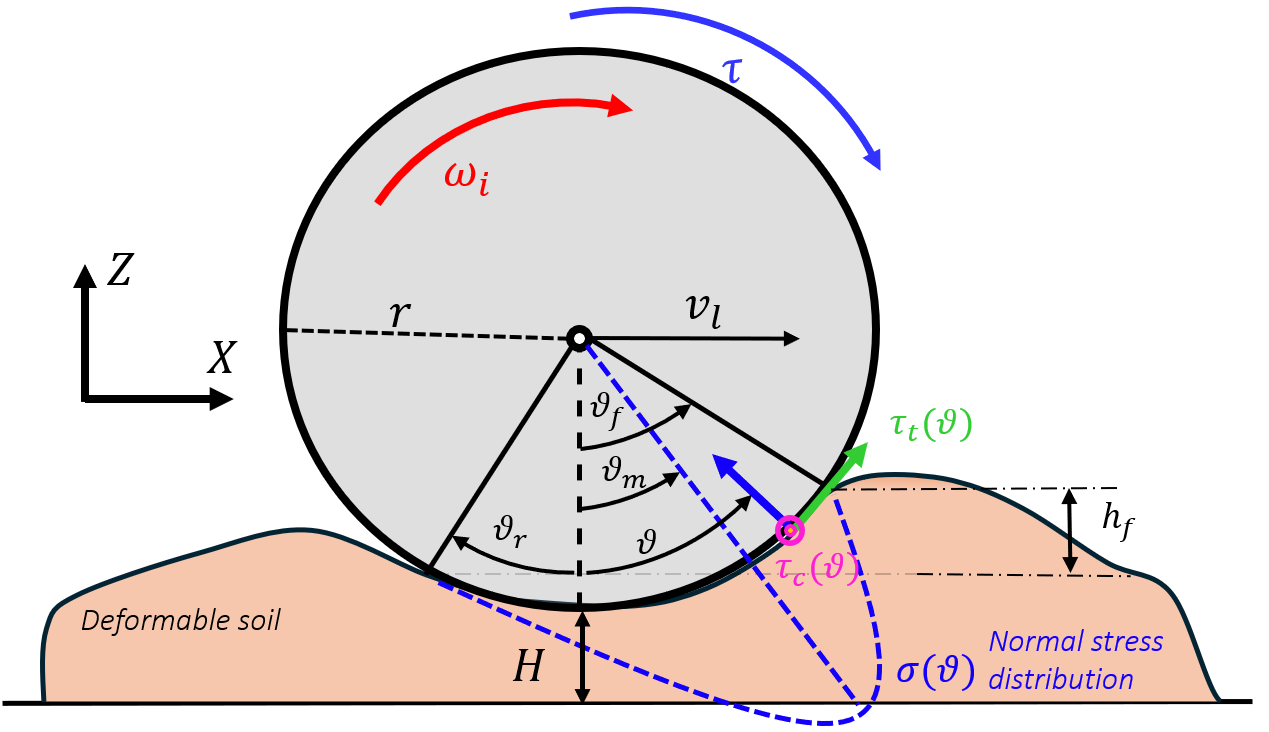}
\caption{Wheel-terrain interaction model} \label{Fig:wheel-terrain}
\end{figure}

Here, $u$ is the longitudinal velocity and $v$ is the lateral velocity of the vehicle's center of mass expressed in the body-fixed frame, and $\dot{\psi}$ is the yaw rate, as shown in Fig.~\ref{Fig:longitudinal}. The global position of the center of mass can be found by integrating the global velocity, which is represented as, 
\begin{align}\label{eq:pos_kinematic}
    \dot{X} = u \cos\psi - v \sin\psi \\
    \dot{Y} = u \sin\psi + v \cos\psi.
\end{align}
The parameters $m$ and $I_z$ represent the vehicle mass and yaw moment of inertia, respectively. The geometric parameters $l_f$ and $l_r$ are the distances between the front and rear wheels from the center of gravity. The control input to these Eqs.~\eqref{Eq:longitudinal_x}--\eqref{Eq:longitudinal_psi} is the steering angle $\delta$, while the wheel forces $F_{lr}, F_{cr}, F_{lf}, F_{cf}$, are computed at each time step using the wheel-terrain interaction (Bekker-Wong-Reece, \cite{wongPredictionRigidWheel1967}) model, illustrated in Fig.~\ref{Fig:wheel-terrain} and explained in section \ref{WTI}. The subscripts $(\cdot)_l$ and $(\cdot)_c$ correspond to longitudinal and cornering (lateral) directions, whereas $(\cdot)_r$ and $(\cdot)_f$ denote the rear and front wheel, throughout this paper. Additional forces included in the bicycle model are the aerodynamic drag forces, which are modeled as, 
\begin{equation}
    f_{ax} = \frac{1}{2}\rho_{air} C_d A_{fx} u^2, ~~~~ f_{ay} = \frac{1}{2}\rho_{air} C_d A_{fy} v^2. 
\end{equation}
Where $\rho_{air}$ is the air density, $C_d$ is the dimensionless drag coefficient, and $A_f$ is the reference area. 

\subsection{Half-car suspension model}
\begin{figure}[h]
        \centering
        \includegraphics[width=0.4\textwidth, keepaspectratio]{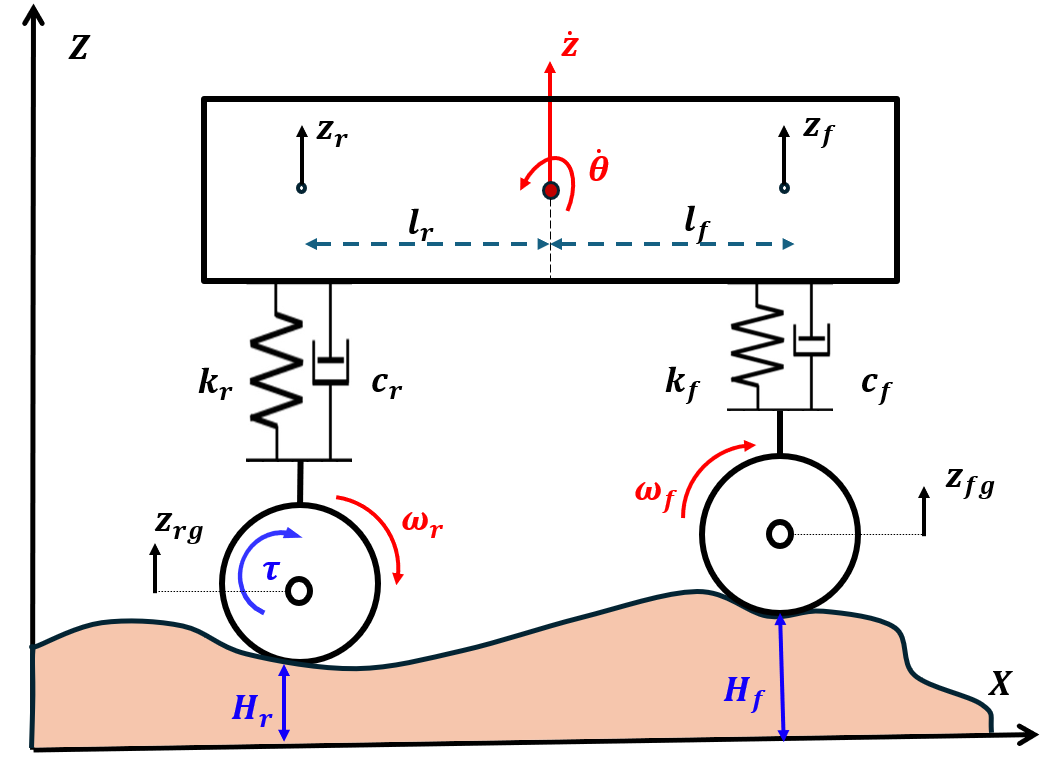}
        \caption{Vertical dynamics using half-car suspension model}\label{Fig:vertical}
\end{figure}

The vertical and pitching dynamics are captured using a half-car model \cite{rajamaniVehicleDynamicsControl2006}, as shown in Fig.~\ref{Fig:vertical}. It includes the sprung mass (vehicle body) supported by the front and rear suspension systems, each modeled as a spring-damper pair connected to the terrain profile under the corresponding rigid wheel. Previously, in \cite{parkInfluenceSoilDeformation2004,BT_acc_2022}, it was demonstrated that incorporating vertical dynamics is crucial for terrain parameter estimation and trajectory prediction accuracy in off-road vehicles operating on deformable terrain. 

The equations of motion are derived by balancing vertical forces and pitching moments, with $z$ and $\theta$ representing the vertical displacement and pitch angle of the vehicle’s center of mass from equilibrium, which are defined with respect to the vehicle’s static equilibrium position and also assumes small deviations in the pitch angle such that  $\sin \theta \approx \theta$ and $\cos \theta \approx 1$. The half-car model is further extended to include the front and rear wheel angular velocities $\omega_i$, as additional states, where $ i=\{f,r\}$ denotes the front and rear wheel subscripts. This formulation enables the computation of wheel slip ratios required for the Bekker-Wong-Reece terramechanics model. The rear wheel is driven by an applied torque ($\tau$), which serves as the primary control input to the half-car model, as detailed in the following equations.
\begin{subequations}\label{eq:half_car}
    \begin{align}\label{eq:half_car_z}
     m\ddot{z} =~& m(v \dot{\psi} \sin{\theta} + u \dot{\theta}) -\sum_{i=f,r} \Big( k_i (z_i - z_{ig}) ~+ \\ & c_i (\dot{z}_i - \dot{z}_{ig}) \Big), \nonumber \\ \label{eq:half_car_theta}
    I_y \ddot{\theta} =&~ k_r (z_r - z_{rg})l_r\cos{\theta} + c_r (\dot{z}_r - \dot{z}_{rg}) l_r\cos{\theta} ~- \\ 
    &  k_f (z_f - z_{fg})l_f\cos{\theta} - c_f (\dot{z}_f \dot{z}_{fg})l_f\cos{\theta}, \nonumber \\
    \dot{\omega}_r =&~ \frac{-r(F_{lr} + f_{rr}) + \tau}{I_{wr} }, \\ \label{eq:half_car_wr}
     \dot{\omega}_f =&~ \frac{-r(F_{lf} + f_{rf})}{I_{wf}}. 
\end{align} \label{eq:half_car_wf}
\end{subequations}
The model includes suspension stiffness $(k_f,k_r)$, damping coefficients $(c_f,c_r)$ and the vertical displacements at the front and rear axles $(z_f,z_r)$, which are related to the center of mass motion as 
\begin{equation}
    z_f = z + l_f \sin \theta,~~ and ~~z_r = z - l_r \sin \theta,
\end{equation} 
which is shown in the Fig.~\ref{Fig:vertical}. The term $z_{gf}$ and $z_{gr}$ represent the vertical positions of the front and the rear wheel centers from the ground contact points, respectively, which are defined by the terrain height profile $H(x,y)$ and the wheel sinkage $(h_s)$ on deformable soil. For a given wheel,
\begin{equation}
    z_{g} = H(x, y) - h_f
\end{equation}
The function $H(x,y)$ is prescribed analytically or generated from terrain data, and it acts as a known input to the system; its value and gradients are assumed to be known. The wheel sinkage is computed from the Bekker–Wong-Reece terramechanics relations described in the next subsection \ref{WTI}. Also, $I_y$ is the pitching moment of inertia, $r$ is the wheel radius and $ f_{r}$ are the rolling resistance forces, which represent the mechanical energy lost as a tire rolls, primarily due to tire deformation and hysteresis. In modeling, it is often expressed as a retarding force, given as
\begin{equation}
    fr = P^{\alpha_R} N^{\beta_R} (A + B v_{l} + C v^2_{l}),
\end{equation}
which captures its dependence on inflation pressure $(P)$, normal load $(N)$, and longitudinal velocity of the tire $(v_{l})$. Here, $\alpha_R$, $\beta_R$ are the tire pressure exponent and $A, B,$ and $C$ are the regression constants. This formulation reflects how rolling resistance increases under heavier loads, lower pressures, and higher speeds, leading to greater energy dissipation as heat in the tire, as discussed in \cite{groverModelingRollingResistance1998,gentPneumaticTire}. 

The normal reaction $N$ at each wheel is computed from the suspension dynamics and the local terrain height. Assuming continuous contact with the terrain profile $z_g = H(x,y)$, the dynamic normal load is expressed as,
\begin{equation}\label{eq:normal_load}
    N = \frac{1}{2}mg -k(z-z_g) - c(\dot{z} - \dot{z}_g) + m_w\ddot{H}(x,y)
\end{equation}
where, $m_w$ is the mass of the wheel. The vehicle parameters used in simulation for generating the datasets are given in Table \ref{tab:vehicle_parametrs}.

\begin{table}[h]
\caption{Vehicle Parameters}\label{tab:vehicle_parametrs}
\setlength{\tabcolsep}{8pt} 
\centering{%
\begin{tabular}{c c}
\toprule
Parameter & Value \\
\midrule
Tire radius $r$ & $0.33~m$ \\
Tire width $b$  &  $0.2286~m$ \\
Sprung mass $m$ & $452~Kg$ \\
Unsprung mass $m_w$ & $30~Kg$ \\
Wheel base $l_f + l_r$ &  $2.719~m$ \\
Suspension stiffness $k_f,k_r$ & $5\times 10^{3}~N/m$ \\
Suspension damping $c_f,c_r$ & $300~Ns/m$ \\
\bottomrule
\end{tabular}
}%
\end{table}

\subsection{Wheel-Terrain Interaction Modeling} \label{WTI}

To evaluate the longitudinal $(F_l)$ and lateral $(F_c)$ forces produced by a wheel on deformable terrain, this work utilizes a terramechanics model from the formulations of Bekker, Wong and Reece, \cite{bekkerOfftheroadLocomotionResearch1960,wongPredictionRigidWheel1967}. It will be assumed that the parameters of these models are known, although these can be estimated real time in-situ \cite{shamraoEstimationTerramechanicsParameters2018, BT_acc_2022}. Soft soil exhibits highly nonlinear effects, such as sinkage and shear deformation, and this model accurately captures those behaviors through analytical pressure-sinkage and shear-stress relations. It determines how the soil is stressed at each angular location along the wheel-terrain contact patch and then integrates the normal pressure $(\sigma)$ and shear stresses $( \tau_t, \tau_c)$ to obtain the total tire forces. It provides sufficient physical detail while remaining computationally practical, making it appropriate for off-road vehicle simulation.

The vertical compression of the soil is effectively approximated using Bekker's pressure-sinkage law,
\begin{equation}\label{eq:pressure_sinkage}
    \sigma (\vartheta) = \Big( \frac{k_c}{b} + k_{\phi}\Big)h(\vartheta)^n,
\end{equation}
where, $b$ is the wheel width, $n$ is the sinkage exponent, $k_c$, $k_{\phi}$ are the cohesive and friction moduli and $\vartheta$ is the contact angle. The local sinkage $h$ is represented as,
\begin{equation}
    h(\vartheta) = \begin{cases}
        ~r (\cos \vartheta - \cos \vartheta_f), ~~~ \vartheta_m  \leq \vartheta \leq  \vartheta_f, \\
       ~r (\cos \vartheta_e - \cos \vartheta_f), ~~ \vartheta_r \leq \vartheta \leq \vartheta_m
    \end{cases}
\end{equation}
where the wheel enters the soil at angle $\vartheta_f$, exits at $\vartheta_r$ and the angle of maximum normal stress is at $\vartheta_m$. These angles depend on the soil parameters $(a_0, a_1, \lambda_r)$ and the slip ratio $s$,
\begin{align}
    \vartheta_f =&~ \cos^{-1}\Big( 1 - \frac{h_f}{r}\Big), \\
    \vartheta_m =&~ \Big(a_0 +a_1 s \Big) \vartheta_f, \\
    \vartheta_r =&~-\cos^{-1}\Big( 1 -\lambda_r\frac{h_f}{r}\Big), \\
    \vartheta_e =&~ \vartheta_f - \Big( \frac{\vartheta - \vartheta_r}{\vartheta_m - \vartheta_r}\Big) (\vartheta_f - \vartheta_m).
\end{align}

In addition to normal stress, the longitudinal or tangential component $\tau_t(\vartheta)$ and lateral or cornering component $\tau_c(\vartheta)$ of the shear stress provide the mechanisms that generate driving, braking, and lateral cornering forces. The Wong-Reece shear model characterizes how this shear stress develops in the soil, due to the soil parameters like cohesion $(c)$, internal friction $(\phi)$, and the soil’s shear deformation modulus $(k_t, k_c)$ as shown below,
\begin{align}\label{eq:tau_x}
    \tau_t(\vartheta) =&~ \Big(c + \sigma(\vartheta)\tan\phi \Big) \Big(1 - e^{-j_t(\vartheta)/k_t} \Big), \\\label{eq:tau_y}
    \tau_c(\vartheta) =&~ \Big(c + \sigma(\vartheta)\tan\phi \Big) \Big(1 - e^{-j_c(\vartheta)/k_c} \Big).
\end{align}
The shear displacements $j_t(\vartheta)$ and $j_l(\vartheta)$ represent how much the soil shears before the wheel surface leaves the contact patch:
\begin{align}
    j_t(\vartheta) =&~ r \Big((\vartheta_f - \vartheta) - (1-s)(\sin \vartheta_f - \sin \vartheta) \Big), \\
    j_c(\vartheta) =&~ r(1-s)(\vartheta_f- \vartheta)\tan \beta
\end{align}
The slip ratio $(s)$ and slip angle $(\beta)$ are defined as,
\begin{align}
    s =&~ \begin{cases}
        ~\Big( 1 - \frac{\omega r}{v_l} \Big), ~~ |\omega r| > |v_l|,~~ \text{driving} \\
        ~\Big( \frac{v_l}{\omega r} - 1 \Big), ~~ |\omega r| < |v_l|,~~ \text{braking} 
    \end{cases} \\
    \beta =&~ \tan^{-1}(\frac{v_c}{v_l}),
\end{align}
where $v_l$ and  $v_c$ are the longitudinal velocity and lateral (cornering) velocity of the wheel, respectively. 

With the normal and shear stress defined as functions of the wheel angle in Eqs.\eqref{eq:pressure_sinkage}, \eqref{eq:tau_x}, and \eqref{eq:tau_y}, the resultant forces are obtained by integrating these stresses over the angular extent of the contact patch. The total forces applied by the soil on the wheel are,
\begin{align}\label{eq:force_intx}
    F_l =&~ \int^{\vartheta_f}_{\vartheta_r} r b \Big(\tau_t(\vartheta) \cos \vartheta - \sigma(\vartheta)\sin \vartheta \Big)d\vartheta, \\\label{eq:force_inty}
    F_c =&~ \int^{\vartheta_f}_{\vartheta_r} r b \tau_c(\vartheta), \\\label{eq:force_intz}
    F_z =&~ \int^{\vartheta_f}_{\vartheta_r} r b \Big(\tau_t(\vartheta) \sin \vartheta + \sigma(\vartheta)\sin \vartheta \Big)d\vartheta.
\end{align}
The maximum sinkage $(h_f)$ is determined by enforcing equilibrium between the vertical force predicted by the terramechanics model and the wheel’s normal reaction computed using Eq. \eqref{eq:normal_load}. This requires solving,
\begin{equation}
    F_z(h_f | \vartheta) - N = 0.
\end{equation}
A Newton–Raphson iteration is used to find the root of this equation. Once $h_f$ is identified, the full normal and shear stress fields are defined, and the longitudinal and lateral forces can be evaluated from the integral expressions in Eqs. \eqref{eq:force_intx}--\eqref{eq:force_intz}. Clay and sandy loam soil parameters are shown in the Table \ref{tab:soil_parametrs}, which are used in the simulation in later sections.

\begin{table}[h]
\caption{Soil Parameters}\label{tab:soil_parametrs}
\setlength{\tabcolsep}{3pt} 
\centering{%
\begin{tabular}{c c c c}
\toprule
Parameter & Unit & Sandy Loam & Clay \\
\midrule
Cohesion modulus $k_c$ & $[kN/m^{n+1}]$ & 5.27 &  13.19\\
Frictional modulus $k_{\phi}$ & $[kN/m^{n+2}]$ & 1515.04 & 692.15 \\
Cohesion $c$ &  [kPa] & 1.72 & 4.14 \\
Friction angle $\phi$ & [rad] & 0.5061 & 0.2269\\
Sinkage exponent $n$  &  & 0.7 & 0.5 \\
Shear deformation  & $[m]$ & 0.025 & 0.01 \\
modulus $k_t, k_c$  \\
$a_0$ & & 0.18 & 0.43 \\
$a_1$ & & 0.32 & 0.32 \\
$\lambda_r$ & & 0.08 & 0.08\\
\bottomrule
\end{tabular}
}%
\end{table}

\section{Koopman Operator Identification Framework}\label{sec:3}
The strong nonlinearities introduced by the wheel-terrain interaction make the dynamics of off-road vehicles difficult to model and control using standard approaches. This motivates the exploration of Koopman-based data-driven methods, which reveal a linear representation of the system by lifting the input–output data into an appropriate space of observables.

\subsection{Koopman Theory and Finite Linear Embeddings}
Consider a discrete-time dynamical system defined as
\begin{equation*}
    x_{t+1} = F(x_t)
\end{equation*}
where $x_t \in \M \subseteq \RR^d$ denotes the state at time $t$ and $F:\M \mapsto \M$ specifies the map of state update. Let $\G$ be the collection of real-valued functions on $\M$, referred to as observables. For any $g \in \G $, the Koopman operator $\K$ is defined through composition with the dynamics,
\begin{equation}
    \K g(x_t) = (g\circ F)(x_t) = g(x_{t+1}).
    \label{eq:koopman_def}
\end{equation}
The expression in \eqref{eq:koopman_def} highlights that the Koopman operator advances observables rather than states, shifting the analysis from the nonlinear state space to a linear evolution in the function space. As a result, the collection of observables evolves according to a linear infinite-dimensional dynamical system. This property underlies the appeal of Koopman-based modeling, as it provides a linear framework for analyzing and approximating nonlinear dynamical systems, see \cite{lasota_1994, mezic_chaos_2012, brunton2021modernkoopmantheorydynamical} for a comprehensive review.

Within this representation, the eigenfunctions $(\Phi(x))$ of $\K$ evolve exactly through scalar multiplication at each time step, and any observable, in principle, can be expanded as a linear combination of eigenfunctions as $g = \sum c_k \phi_k(x)$. However, neither the Koopman operator nor its eigenfunctions are known, which means the appropriate observable space is also unknown. This creates a central difficulty in Koopman analysis. In practice, we construct a finite-dimensional approximation that requires a set of observables that mimic the action of the unknown eigenfunctions, yet identifying such observables would normally require knowing the operator itself. To overcome this, Koopman methods introduce a finite set of basis functions $\Psi(x) = [\psi_1(x), \psi_2(x), \cdot\cdot\cdot \psi_r(x)] \in \RR^r$  that act as surrogate observables. These basis functions approximate the span of the dominant eigenfunctions so that the action of the infinite-dimensional operator can be projected onto a finite lifted space. Then, these basis function evolves through an approximation of the Koopman operator $K_{\Psi}$ as 
\begin{equation}
    \K g(x_t) \approx \K [c^{\intercal} \Psi](x_t) \approx (K_{\Psi}c)^{\intercal} \Psi(x_t) = c^{\intercal} \Psi(x_{t+1}).
    \label{eq:koopman_proj}
\end{equation}
Some of the approaches to selecting basis functions include Dynamic Mode Decomposition (DMD), which utilizes state variables as the basis (see \cite{H_Tu_2014}), Extended DMD, which employs hand-crafted dictionaries (see \cite{williams2015edmd,williams2016edmdc,korda_mezic_2018}), and learn the basis jointly with the operator through neural networks or kernels see \cite{kevrekidis_chaos_2017,lusch2018deep,Williams2016AKM}. When control inputs are present, selecting an appropriate dictionary becomes even more challenging because the lifted dynamics must now satisfy a controlled linear evolution. However, designing or learning such a dictionary is only one part of the problem; the other problem is to find the finite approximation of the Koopman operator.

For a discrete nonlinear controlled system of the form,
\begin{equation}
    x_{t+1} = F_u(x_t, u_t),  ~~~~ y_t = g(x_t)
\end{equation}
where $u_t \in \mathcal{U} \subset \RR^m$ is the input and  $F_u : \M \times \U \mapsto \M $ is the state transition function, see \cite{proctor2018_dmdc}. It is also assumed that the measurements of the observable function are available and are the outputs $(y_t)$ of the system. Another assumption in the finite-dimensional representation for the controlled case is the choice of a linear time-invariant model of the form,
\begin{subequations}\label{eq:linear}
\begin{align}
z_{t+1}  & = A z_t + Bu_t,  \\
y_t & = C z_t,
\end{align}
\end{subequations}
which is commonly adopted for practical reasons against the bilinear form, particularly when the goal is to approximate the dynamics for simpler estimation and control design. With these assumptions, the Koopman learning problem for the controlled system can be expressed as a coupled optimization over the latent linear dynamics and lifting map,
\begin{align}
    \min_{A,B,C,z,\Psi} \sum_t \Big( ||y_t - C z_t ||^2 +||z_{t+1} -A z_t - B z_t ||^2  \nonumber \\ + ~||z_t - \Psi(x_t) ||^2 \Big)
\end{align}
As shown in \cite{lianLearningFeatureMaps2019}, the problem can be separated into two independent subproblems without loss of generality. The first subproblem involves optimizing the first two terms, which removes the dependence on the unknown dictionary and identifies only the latent linear dynamics and latent state sequence. 
Subspace identification (SSID) methods are ideally suited to solve this problem because they construct input-output Hankel matrices, isolate the extended observability subspace, determine the system order, and reconstruct the latent state sequence $z_t \in \RR^r$. In doing so, SSID provides a data-driven estimate of the finite-dimensional Koopman dynamics without requiring any prior specification of observables.  

The second subproblem learns the lifting map via supervised regression,
\begin{equation}\label{eq:lifted_mapping}
    \min_{\Psi} \sum_t ||z_t - \Psi(x_t) ||^2.
\end{equation}
Although Gaussian processes are used here to learn the lifting map $\Psi(x_t)$, they are only one possible choice for solving the second subproblem. Any expressive regressor, such as neural networks or alternative kernel methods, can be used to map the physical state $x_t$ to the latent Koopman coordinates $z_t$. This lifting step is essential because the latent states obtained from subspace identification live in an abstract coordinate system determined by the SVD of the projected Hankel matrices. Learning $\Psi(x_t)$ ensures that the physical state is consistently mapped into this Koopman subspace by optimizing on Eq. \eqref{eq:lifted_mapping}. Delay-coordinate embeddings, as seen in \cite{Brunton_2017}, are useful for reconstructing past dynamics but cannot provide a consistent and general state-to-latent mapping. They rely on requiring several past measurements, do not align with the SSID-identified coordinate system, and therefore cannot initialize or update the Koopman model for prediction or control. A learned lifting map is thus necessary to connect the true state space to the latent linear dynamics uncovered by subspace identification.

\subsection{Brief Review on Subspace Identification Methods}
In this subsection, we summarize how projection-based subspace identification techniques are used to identify the linear time-invariant system in Eq. \eqref{eq:linear} when multiple independent data sets are available in the form of input-output sequences $\D^i = \{u^i_t, y^i_t\}^n_{t=0}, ~~i =1, \cdots, N$. Where $i$ corresponds to a separate experiment and $t$ denotes the sampling instant within the experiment. We assume that the system $(A, C)$ is observable, but make no direct assumption about the order $r$ of the system. However, there is a maximum cap on $r$ when we choose the parameter $l$ for the depth of the Hankel matrix. The Hankel matrices for each data record consisting of $n+1$ measured $I/O$ data can be described with depth $l$ and length $s = n-l+1$ with a condition to have sufficient columns such that $s > lm+r$, see \cite{holcombSubspaceIdentificationMultiple2017}.
\setlength\arraycolsep{1pt}
\begin{equation}
    \begin{split}\label{eq:block_Hankel}
        ^iY_l = 
        \begin{pmatrix}
            ^iy_{t_0} & ^iy_{t_1} & \cdots & ^iy_{t_{n-l}}\\
            ^iy_{t_1} & ^iy_{t_2} & \cdots & ^iy_{t_{n-l+1}}\\
            \vdots  & \vdots    & \ddots & \vdots\\
            ^iy_{t_{l-1}} & ^iy_{t_l} & \cdots & ^iy_{t_n}
        \end{pmatrix}, \\
         ^iU_l = 
        \begin{pmatrix}
            ^iu_{t_1} & ^iu_{t_2} & \cdots & ^iu_{t_{n-l+1}}\\
            ^iu_{t_2} & ^iu_{t_3} & \cdots & ^iu_{t_{n-l+2}}\\
            \vdots  & \vdots    & \ddots & \vdots\\
            ^iu_{t_{l-1}} & ^iu_{t_{l}} & \cdots & ^iu_{t_{n}}
        \end{pmatrix}.
    \end{split}
\end{equation}
This can also be extended for multiple data records by simply appending the new $I/O$
data Hankel matrices to the previous one to form a mosaic- Hankel matrix as, 
\begin{equation}\label{eq:mosaic-Hankel}
\begin{split}
      \mathbb{Y}_N &= ~^{1:N}Y_l = \begin{bmatrix}
        ^1Y_l, & \cdots, & ^NY_l
    \end{bmatrix}    \in \RR^{pl \times Ns}, \\
     \mathbb{U}_N &= ~^{1:N}U_l = \begin{bmatrix}
    ^1U_l, & \cdots, & ^NU_l
    \end{bmatrix} \in \RR^{m(l-1) \times Ns}, \\
    \hat{\mathbb{Z}}_{0,N} &= ~\hat{Z}_{0,1:N} = \begin{bmatrix} \hat{Z}_{0,1}, \hat{Z}_{0,2}, \cdots, \hat{Z}_{0,N} \end{bmatrix} \in \RR^{r \times Ns}.
\end{split} 
\end{equation}

From Eq.\eqref{eq:linear} it follows,
\begin{equation}\label{eq:concat_multi}
    \mathbb{Y}_N = \Gamma_l \hat{\mathbb{Z}}_{0,N} + \H_l \mathbf{U}_N.
\end{equation}
where,
\begin{align}
    \begin{split}
        \Gamma_l = \begin{bmatrix}
            C \\
            CK \\
            CK^2\\
            \vdots \\
            CK^{l-1}
        \end{bmatrix},~~~
        \H_l = \begin{bmatrix}
            D & 0 & 0 & \cdots & 0 \\
            CB & D & 0 & \cdots & 0 \\
            CKB & CB & D & \cdots & 0 \\
            \vdots & \vdots & \vdots & \ddots & \vdots \\
            CK^{l-2}B & \cdot & \cdot & \cdots & D
        \end{bmatrix}.
    \end{split}
\end{align}
where $\Gamma_l \in \RR^{lp\times r}$ is the extended observability matrix, $\H_l \in \RR^{lp \times lm}$ is the lower triangular block-Toeplitz matrix and  $\hat{Z}_{0,i} \in \RR^{r\times s}$ is the realization of the lifted state that corresponds to the initial condition of each column trajectory for the dataset $i$, in the original state space. To isolate the portion of the data determined by $\Gamma_l$, the input-driven term is removed by projecting onto the orthogonal complement of the row space of $\mathbb{U_N}$ defined as 
$\Pi_{\mathbb{U}_N}^{\perp} = I - \mathbb{U}_N^{\intercal}(\mathbb{U}_N \mathbb{U}_N^{\intercal})^{-1}\mathbb{U}_N$ which gives us,
\begin{equation}\label{eq:concat_without_u}
    \mathbb{Y}_N \Pi_{\mathbb{U}_N}^{\perp} = \Gamma_l \mathbb{Z}_{0,N} \Pi_{\mathbb{U}_N}^{\perp},
\end{equation}
whose left-hand side is known, and we call it a projected data matrix. The column space of the projected data matrix coincides with that of the extended observability matrix $\Gamma_l$. Hence extracting the column space of $\Gamma_l$ using SVD,
    \begin{equation}\label{eq:svd}
        \mathbb{Y}_N \Pi_{\mathbb{U}_N}^{\perp}  = \begin{bmatrix} Q_r & Q_{l-r} \end{bmatrix} \begin{bmatrix}\Sigma_r & 0 \\ 0 & \Sigma_{l-r} \end{bmatrix} \begin{bmatrix}
            V_r^{\intercal} \\ V_{l-r}^{\intercal} \end{bmatrix} \implies \hat{\Gamma}_r = Q_r \Sigma_r^{\frac{1}{2}}
\end{equation}
As $\Sigma_r$ contains the $r$ largest singular values, $\Gamma_l \approx \hat{\Gamma}_r$ is approximated, which determines the order of the lifted dynamics to be $r$. Then $A$ and $C$ matrices can be estimated using $\hat{\Gamma}_r$, and finally, $B$ and lifted state realization $\mathbb{Z}_{0,N}$ can be computed by
solving a least squares problem. For a detailed procedure to compute system matrices and several other subspace-based system identification methods, please refer \cite{overschee_1994,QIN20061502}. 

When the collection of data becomes too large to compute the SVD of the projected Hankel matrix, or when a batch-style update of the model is desired, the identification procedure streams the data and updates the observability subspace recursively. Rather than working directly with the projected matrix $\mathbb{Y}_N \Pi_{\mathbb{U}_N}^{\perp}$, the algorithm operates on its square symmetric form called the compressed data matrix $\Xi_N$,
\begin{equation}
    \Xi_N = \mathbf{Y}_N \Pi_{\mathbf{U}_N}^{\perp}\mathbf{Y}^{\intercal}_N.
\end{equation}
This representation of the data matrix facilitates recursive updates while preserving the subspace that helps determine the system matrices. For a comprehensive derivation of Eqs. \eqref{eq:R_SSID1}–\eqref{eq:R_SSID5}, which is based on the matrix inversion lemma, is discussed in \cite{Oku1999AR4}. This source illustrates that the recursive update equations are an extension of the recursive least squares method. The PO-MOESP version of the subspace identification algorithm introduced by Verhaegen and Dewilde in  \cite{Verhaegen1992SubspaceMI} can be used to treat both process and measurement noise in the system, which was further extended by Oku and Kimura in \cite{Oku1999AR4} by providing a foundation for the recursive update of $\hat{\Gamma}_r$. In Loya et al. \cite{LBT_dsc_2023, LT_mecc_2024}, a comprehensive treatment of multiple data records for recursive subspace identification with efficient updates, as well as its connections to Koopman-based modeling, is provided (see Algorithm \ref{alg:R_SSID_GrD}).

\begin{algorithm}[t]
\caption{Recursive SSID with Grassmannian-Guided Data Selection}\label{alg:R_SSID_GrD}
\begin{algorithmic}[1]
\State \textbf{Initialize:} $ \Xi_i:= \mathbf{Y}_i \Pi_{\mathbf{U}_i}^{\perp} \mathbf{Y}_i~,~ P_i:= (\mathbf{U}_i \mathbf{U}_{i}^{\intercal})^{-1}$ and $\mathbf{Y}_i \mathbf{U}_i^{\intercal}$
\For{$i = 0,1 \dots, N$}
     \State \textbf{Sample:} $\D_{i+1} =  \{u_t^{i+1} , y_t^{i+1}\}_{t=0}^n $ \Comment{New Dataset}
     \State \textbf{Construct:} $Y_{i+1}, U_{i+1}$  \Comment{As in Eq.\eqref{eq:block_Hankel} s.t.  $s > lm + r$}
     \State $\Pi_{U_{i+1}}^{\perp} \gets I - U_{i+1}^{\intercal} \Big(U_{i+1} U_{i+1}^{\intercal} \Big)^{-1} U_{i+1}$
     \State $[q,\sigma^2,v^{\intercal}] \gets svd\big(Y_{i+1} \Pi_{U_{i+1}}^{\perp} Y_{i+1}^{\intercal} \big)$
     \State $\hat{\Gamma}_{i+1} \gets q_r \sigma_r^{\frac{1}{2}}$
     \State $G \gets d_{Gr}(\mathbf{\Gamma}_i, \hat{\Gamma}_{i+1})$
     \If{$ G >  \epsilon $}
        \For{$k = 1,2,\dots, s$}
            \State $\mathbf{u}_k \gets U_{i+1}(:,k)$ \Comment{Matlab notation $(:,k)$}
            \State $\mathbf{y}_k \gets Y_{i+1}(:,k)$
            \State \textbf{Update:} 
            \begin{subequations}\label{eq:R_SSID}
                \begin{align}
                & \alpha_{i+1} \gets \big(1+\mathbf{u}_k^{\intercal} P_i \mathbf{u}_k \big)^{-1} \label{eq:R_SSID1} \\
                & e_{i+1} \gets  \mathbf{y}_k - \mathbf{Y}_i \mathbf{U}_{i}^{\intercal} P_i \mathbf{u}_k \label{eq:R_SSID2} \\
                & \Xi_{i+1} \gets  \Xi_i + \alpha_{i+1}~ e_{i+1} e_{i+1}^{\intercal} \label{eq:R_SSID3} \\
                & P_{i+1} \gets  P_i - \alpha_{i+1} P_i \mathbf{u}_k           \mathbf{u}_k^{\intercal} P_i \label{eq:R_SSID4} \\
                & \mathbf{Y}_{i+1} \mathbf{U}_{i+1}^{\intercal} \gets  \mathbf{Y}_i \mathbf{U}_i^{\intercal} + \mathbf{y}_k \mathbf{u}_k^{\intercal} \label{eq:R_SSID5}
            \end{align}        
            \end{subequations} 
        \EndFor
        \State $[Q,\Sigma^2,V^{\intercal}] = svd(\Xi_{i+1})$
        \State $\Gamma_{i+1} = Q_r \Sigma_r^{\frac{1}{2}}$ \Comment{'r' dominant singular values}
    \EndIf  
\EndFor
\State \textbf{Obtain:} $(A,C,B,\hat{\mathbb{Z}}_{0,N})$  
\begin{subequations}\label{eq:system_matrices} \begin{align}\label{eq:system_matrices_a}
    C =&~ \mathbf{\Gamma}_N(1:p,:),\\ \label{eq:system_matrices_b}
    A =&~ \Big( \mathbf{\Gamma}_N(1:(l-1)p,:) \Big)^{\dagger}~\mathbf{\Gamma}_N(p+1:lp,:),\\  \label{eq:system_matrices_c}
    B, \mathbb{Z}_{0,N} =&~ \arg \min_{B,\mathbb{Z}_{0,N}} \Big( ||\mathbb{Y}_N - \Gamma_l \mathbb{Z}_{0,N} + \mathcal{H}_l \mathbb{U}_N ||^2\Big).
\end{align} 
\end{subequations}

\State \textbf{Lifting:} $z_j|\D_{GP} \sim \Psi(x_t) =  \mathcal{GP}\big(\mu_{z_j| \D_{GP}}, k_{z_j | \D_{GP}} \big)_{j=1}^r$
\begin{equation}\label{eq:GP}
    \min_{\Psi} \sum_t ||\hat{Z}_t - \Psi(x_t) ||^2.
\end{equation} 
\end{algorithmic}
\end{algorithm}

After the update of the data matrix $\Xi_i$ indirectly updates the extended observability matrix using SVD and determines the system order based on 'r' dominant singular values. This enables us to update the Koopman operator and, when necessary, adjust the system order. The corresponding system matrices $(A, B, C)$ and latent state realization $Z_{0}$ then can be computed by solving the equations \eqref{eq:system_matrices_a}-\eqref{eq:system_matrices_c}. For equation \eqref{eq:system_matrices_c}, we use gradient descent optimization. The underlying principle of subspace-based identification is that the geometry of the projected data matrices encodes the essential system dynamics. This observation also makes it possible to quantify how similar two datasets are by comparing the subspaces they generate. Consequently, a Grassmannian distance can be used to filter out redundant datasets and limit recursive updates only to experiments that introduce genuinely new dynamical information. After each recursive update step using Eqs. \eqref{eq:R_SSID1}–\eqref{eq:R_SSID5}, we compute the Grassmannian distance between the accumulated observability subspace $\mathbf{\Gamma}_i$, obatained from all datasets $\mathcal{D}_{1:i}$, and the subspace $\Gamma_{i+1}$, extracted from the new dataset $\mathcal{D}_{i+1}$. A threshold $\epsilon > 0$ is selected such that $G < \epsilon$ designates the incoming dataset as $\epsilon$-redundant, in which case it is omitted from the recursion and if $G > \epsilon$ the dataset is deemed informative and is incorporated into the update. The Grassmannian $\mathrm{Gr}(k,n)$ represents all possible $k$-dimensional linear subspaces of an $n$-dimensional vector space, and the Grassmannian distance is a mathematical measure of the dissimilarity between the two subspaces in a high-dimensional vector space. Then for subspaces $\mathbf{\Gamma_1, \Gamma_2} \in \mathrm{Gr}(r,lp)$, we form matrices $\Gamma_1,\Gamma_2 \in \RR^{lp \times r}$ whose columns are their respective orthonormal bases; then,
\begin{equation}\label{eq:Grass_dist}
   G = d_{\mathrm{Gr}(r,lp)}(\mathbf{\Gamma}_1, \mathbf{\Gamma}_2) = \Big[ \sum_{k=1}^r \theta_k^2 \Big]^{1/2}
\end{equation}
where, $\theta_k=cos^{-1}\big(\sigma_k(\Gamma_1^{\intercal}\Gamma_2)\big)$ is the principal angle between $\Gamma_1$ and $\Gamma_2$ and $\sigma_k$ gives the $k^{th}$ singular values of the matrix $\Gamma_1^{\intercal}\Gamma_2$. The Grassmannian distance for equidimensional subspace can be extended to subspaces of different dimensions as shown in \cite{ye2016schubert}. Therefore, even when system order $r$ is changed in algorithm \ref{alg:R_SSID_GrD}, the Grassmannian distance can still be used to differentiate between the identified subspaces.  

\section{Simulation and Results}\label{sec:4}

\subsection{Vehicle simulation dataset and training}
A high-fidelity 7 degrees of freedom vehicle model, see Eqs.\eqref{Eq:longitudinal} and Eqs.\eqref{eq:half_car}, with Bekker soft-soil dynamics, see section \ref{WTI}, was used to generate a large dataset of off-road vehicle trajectories with two soils, 'clay' and 'sandy loam'. Each trajectory lasted 20 seconds at 100 Hz, yielding 2001 samples. A total of 1600 trajectories were generated for each soil with randomized initial yaw angle, forward speed, and wheel velocities to ensure state diversity. The dataset was collected using Clemson University's Palmetto cluster \cite{antao2024modernizing}. 
Steering inputs were drawn from a set of excitation patterns that included straight driving, multisine slalom, fishhook maneuvers, constant-radius (“circle”) steering, and broadband multisine signals, with amplitudes and frequencies varied within safe limits and steering saturated at $\pm 0.35$ rad. Torque excitation was independently selected from ramp profiles, multisine inputs, pseudo-random binary sequences (PRBS), and linear chirp signals, with all torque commands constrained to the range $[0, 0.95 T_{max}]$, where $T_{max} = 130N$. Amplitudes, tone selections, slopes, dwell times, and dither levels were randomized within predefined bounds. Disjoint multisine frequency pools and a small stochastic overlay ensured persistence of excitation, which was verified using the Hankel-rank test. The soil parameters used for the Bekker model are summarized in Table \ref{tab:soil_parametrs}, and all vehicle mechanical and geometric properties are provided in Table \ref{tab:vehicle_parametrs}. Each trajectory was simulated in MATLAB using ODE113 with event handling where simulations were terminated to capture discontinuities such as wheel lift-off and also when the vehicle speed approached zero, as near-stationary conditions led to numerical difficulties in integrating the dynamics. 
The Koopman model was trained using Algorithm~\ref{alg:R_SSID_GrD}, with the measured outputs chosen as the partial state vector $[u,v,\dot{\psi}]$ and the inputs as $[\delta \,\tau]$. A sampling time of $\Delta t = 0.01\,$s was used, and the Hankel row size was set to $400$. The Gaussian process lifting functions and their hyperparameters were learned using MATLAB’s fitgpr, and the parameters in Eq.~\eqref{eq:system_matrices_c} were estimated via gradient descent (Adam). All training data reported here were generated on flat terrain. 

\subsection{Model Prediction Performance}

Following the dataset construction described in the previous section, we trained two separate Koopman models for each soil using data generated with the corresponding terramechanics parameters because clay and sandy loam soil exhibit substantially different vehicle–terrain dynamics. 

\begin{figure}[t]
     \begin{subfigure}[b]{0.24\textwidth}
         \centering
         \includegraphics[width=\textwidth]{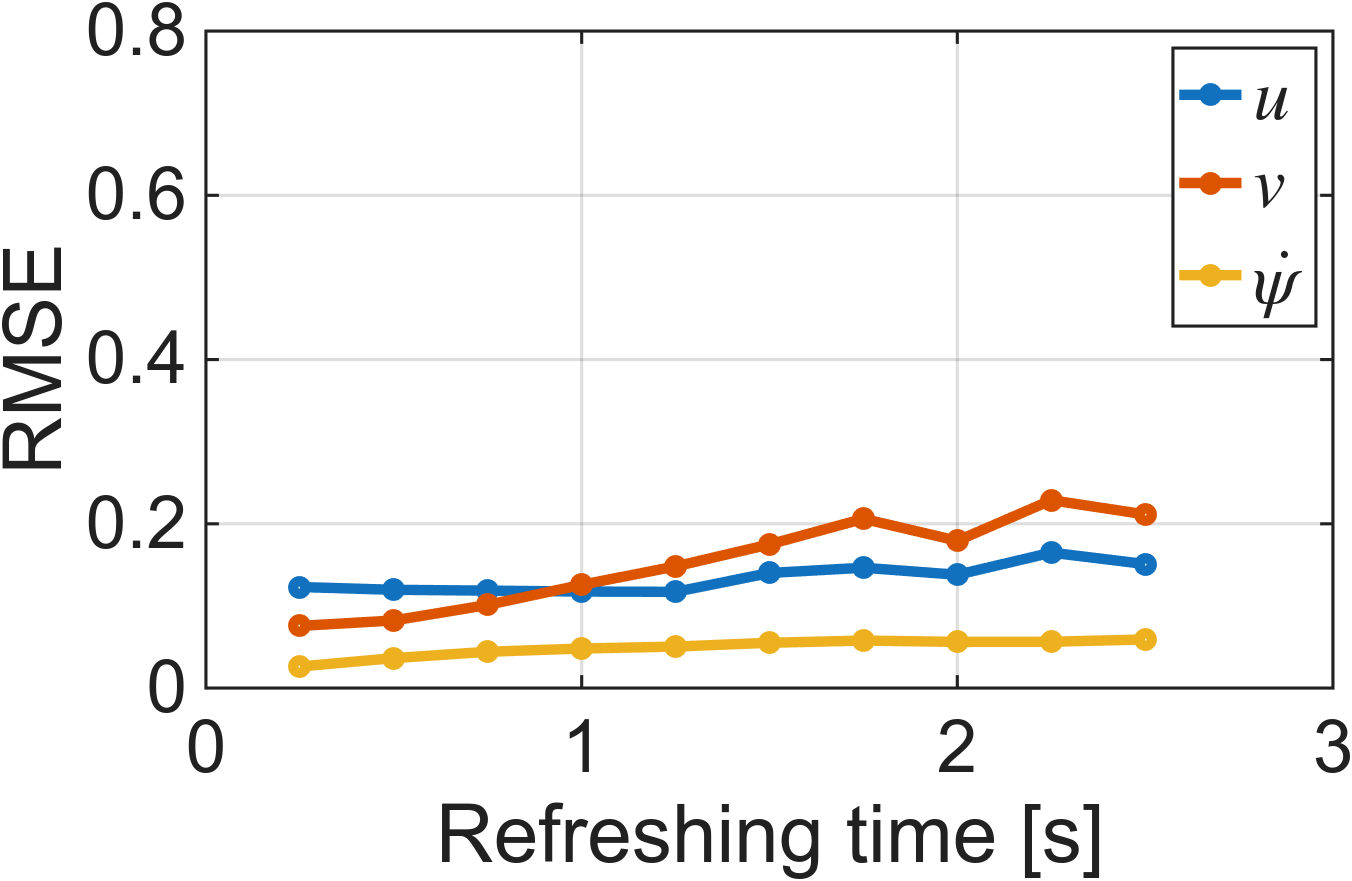}
         \caption{}
         \label{fig:SL_refresh}
     \end{subfigure}
     \begin{subfigure}[b]{0.24\textwidth}
         \centering
         \includegraphics[width=\textwidth]{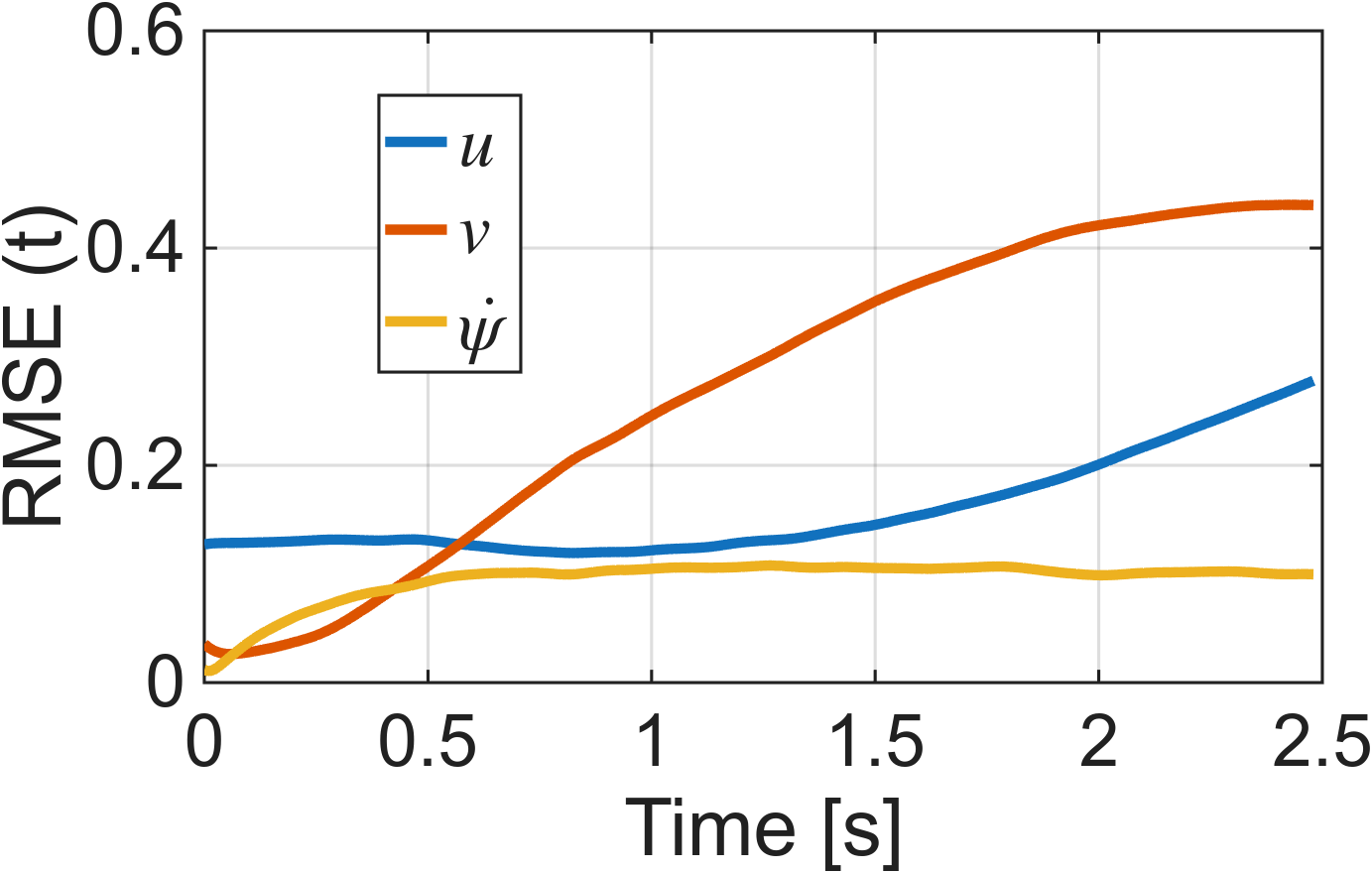}
         \caption{}
         \label{fig:SL_err_time}
     \end{subfigure}
          \begin{subfigure}[b]{0.24\textwidth}
         \centering
         \includegraphics[width=\textwidth]{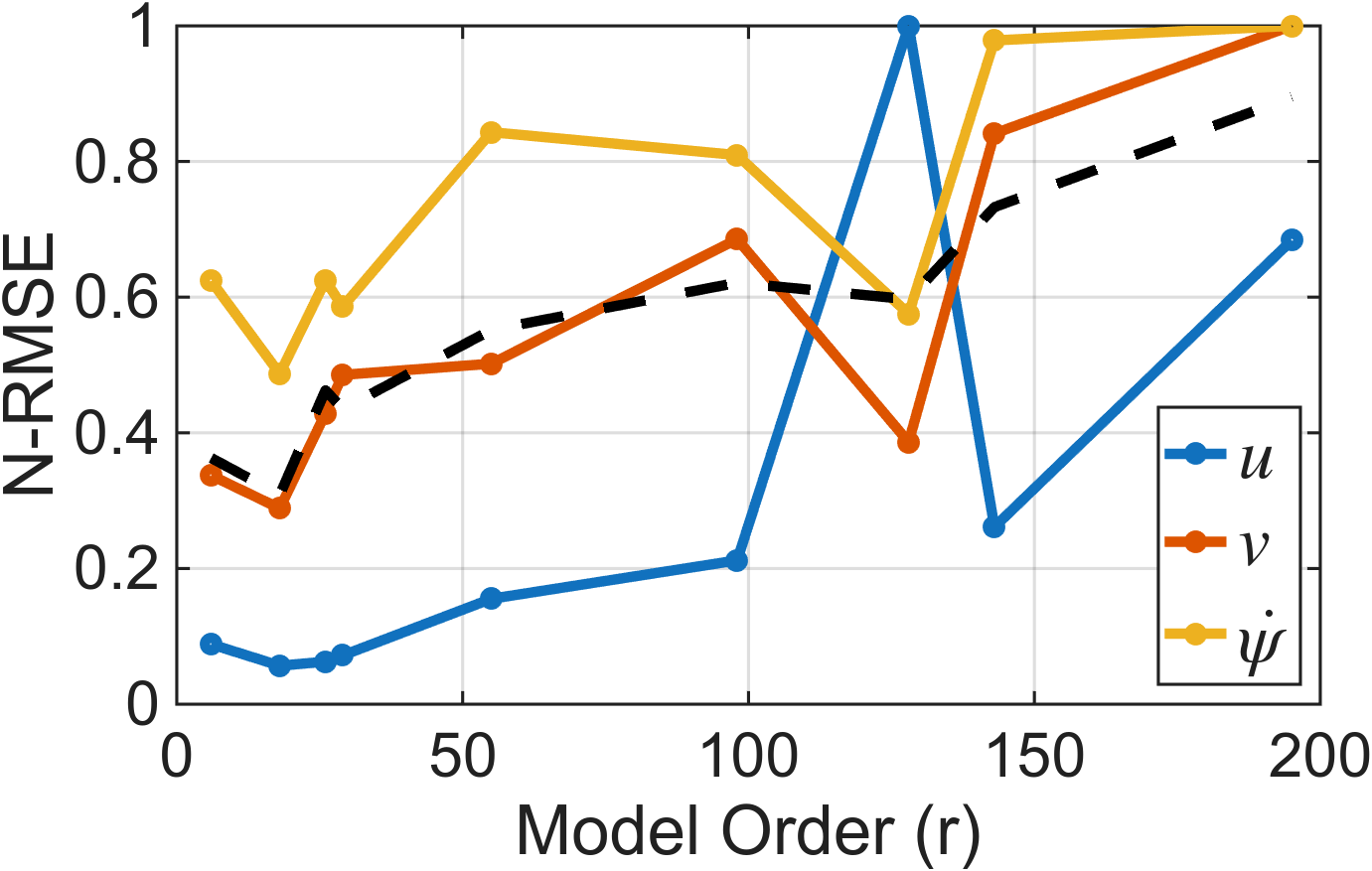}
         \caption{}
         \label{fig:SL_MC}
     \end{subfigure}
          \begin{subfigure}[b]{0.24\textwidth}
         \centering
         \includegraphics[width=\textwidth]{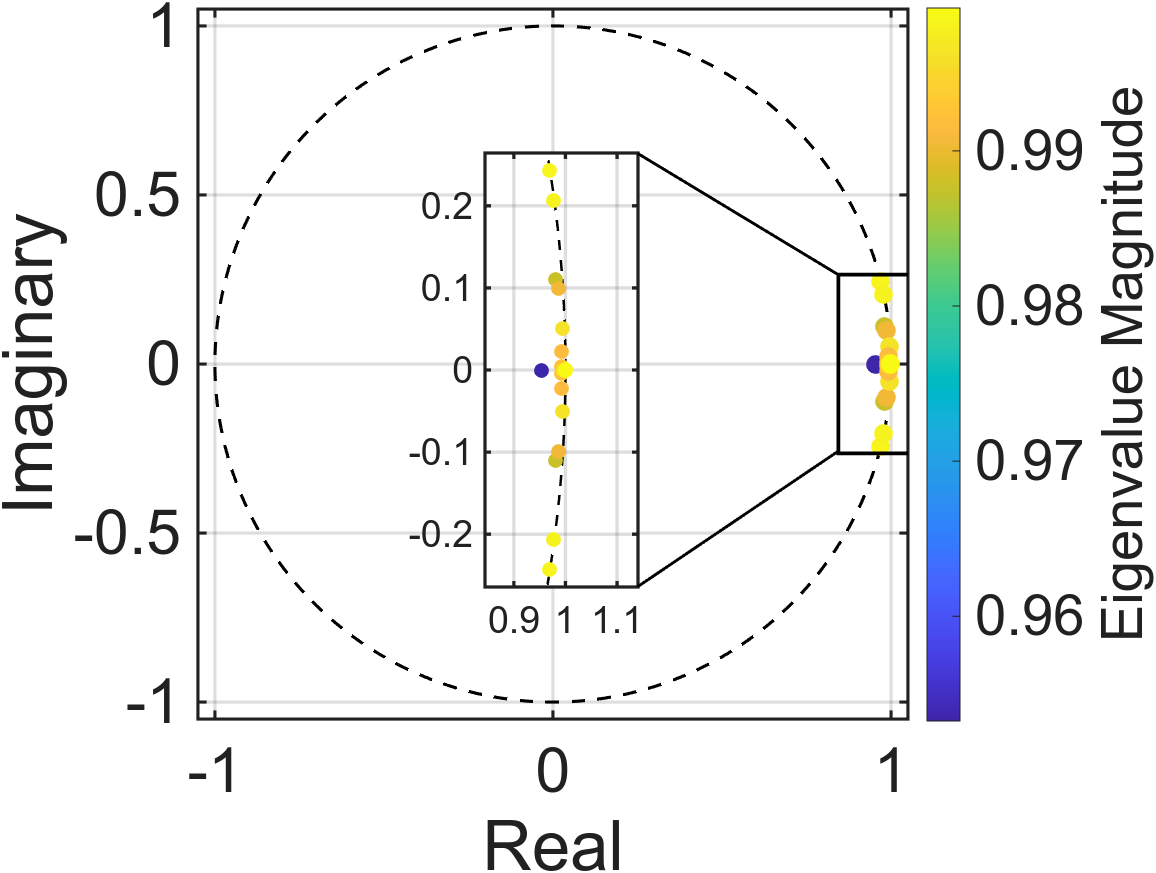}
         \caption{}
         \label{fig:SL_eigen}
     \end{subfigure}
     \caption{Sandy loam soil results for Koopman model selection and prediction: (a) RMSE versus refresh time, (b) open-loop RMSE growth with time, (c) normalized RMSE versus model order $r$ (dotted line: mean across outputs), and (d) eigenvalue spectrum of the identified Koopman $A$ matrix (unit circle shown for reference).}\label{fig:sandyloam_results}
\end{figure}

\begin{figure}[t]
        \centering
        \includegraphics[width=0.48\textwidth, keepaspectratio]{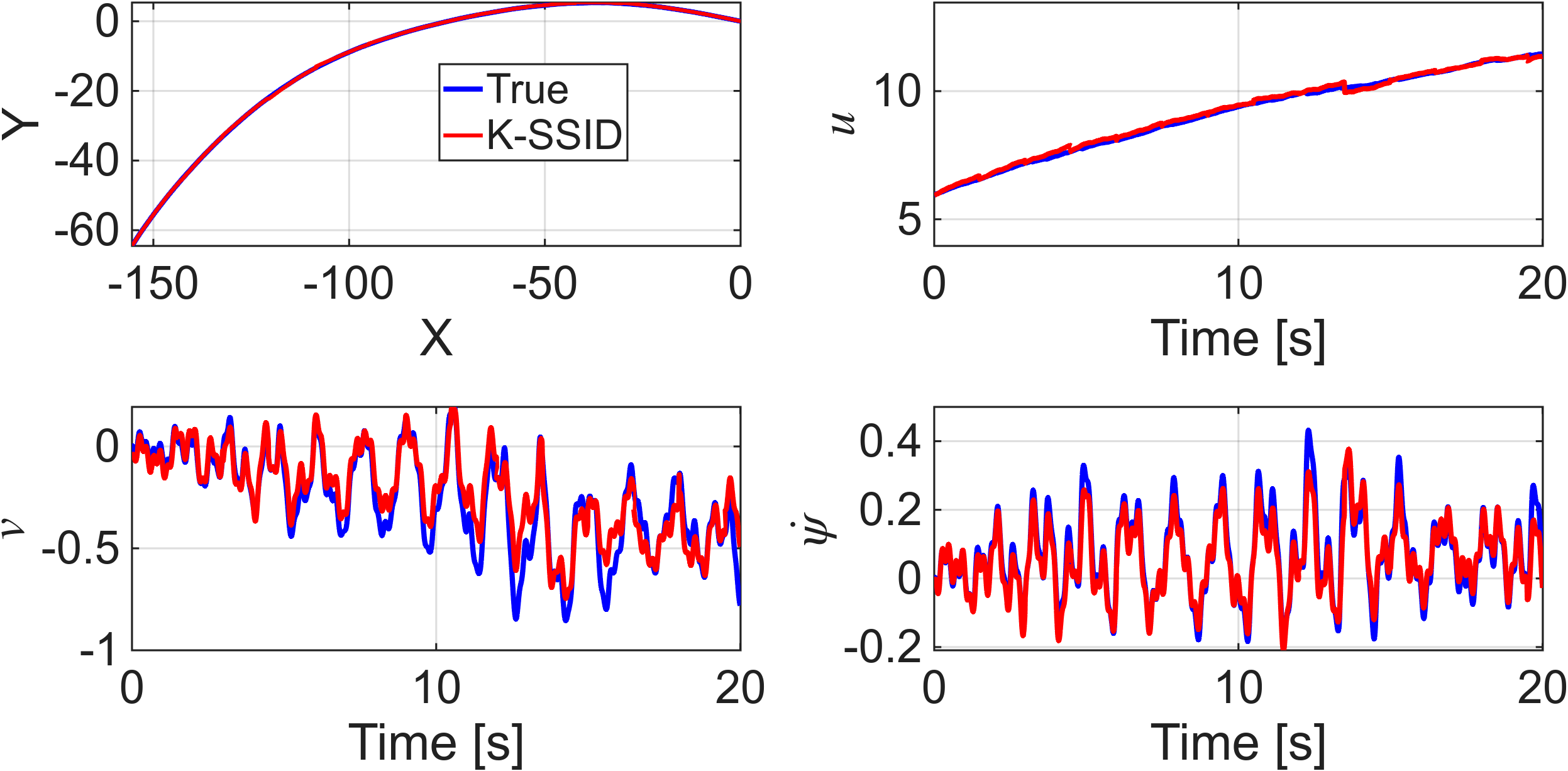}
        \caption{Sample trajectory prediction on sandy loam soil.}
        \label{fig:SL_traj_pred}
\end{figure}

For each soil-specific model, the Koopman model order $r$ was selected from a model-order sweep by varying the singular-value cut-off range used to retain the most informative modes in Eq.~\eqref{eq:svd}. For each resulting model root mean squared error (RMSE) is calculated. For each output prediction $\hat{y}$ and true measured output $y =[u, v, \dot{\psi}]$ RMSE is given as,
\begin{equation}
RMSE_y =  \sqrt{\frac{1}{N_t}\sum_{i=1}^{N_t} \Bigg( \frac{1}{n} \sum_{t=1}^{n}(\hat{y}_{t} - y_{t})^2}\Bigg) ,  
\end{equation}
The RMSE was computed for over $N_t = 100$ test trajectories with each trajectory of $20 s$, that is, $n=2000$ time steps. This error was generated using the same soil parameter set as the model under evaluation and using a $2.5 s$ refresh time, meaning the predictor was periodically re-initialized every $2.5 s$ using the true measured outputs before rolling out the next segment. The RMSE was normalized by the maximum RMSE across the compared models as,
\begin{equation}
N\text{-}RMSE_{y}(r)
= \frac{RMSE_{y}(r)}
{\max\limits_{r \in \mathcal{R}} RMSE_{y}(r)},
\end{equation}

\begin{figure}[t]
     \begin{subfigure}[b]{0.24\textwidth}
         \centering
         \includegraphics[width=\textwidth]{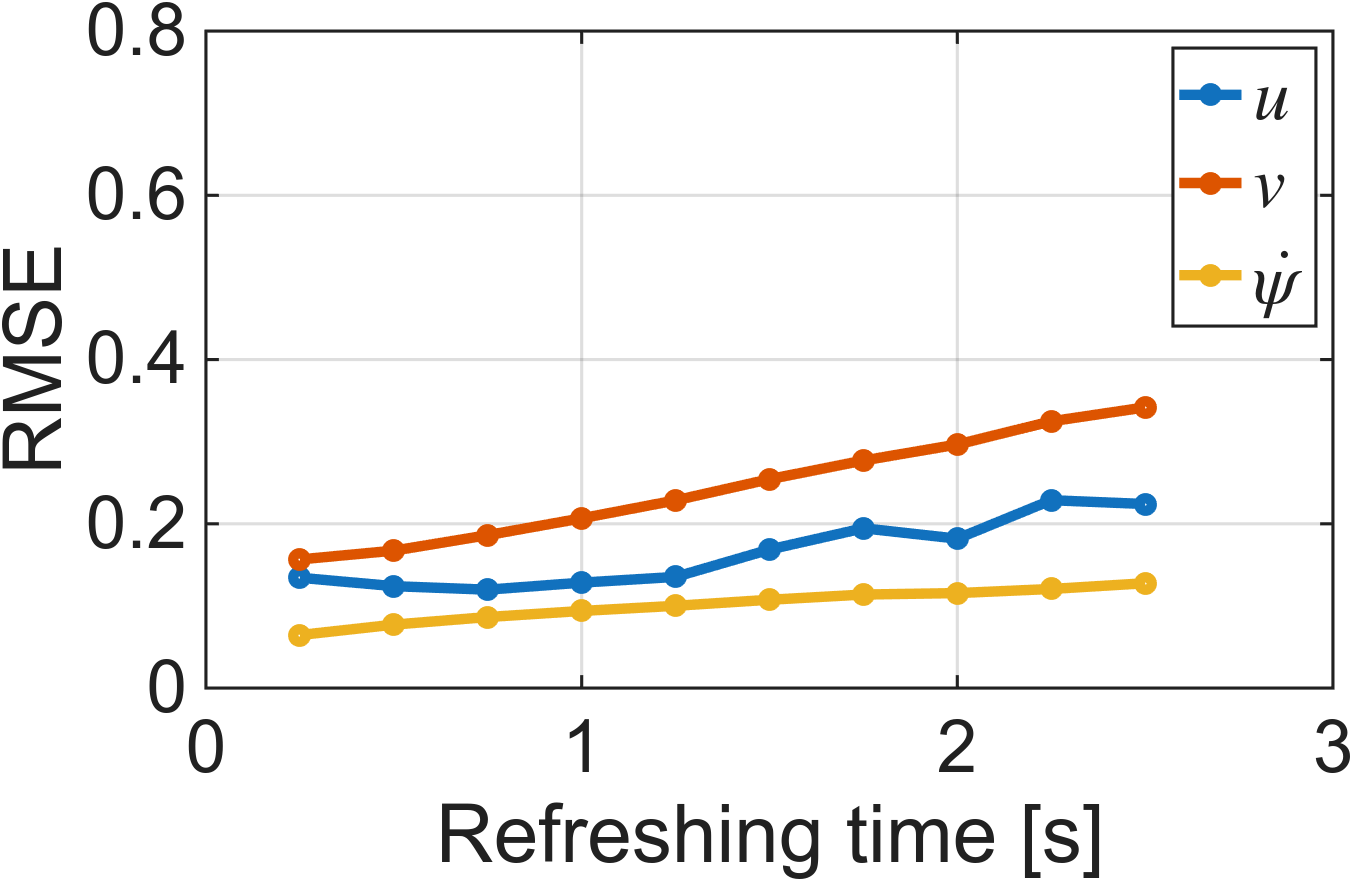}
         \caption{}
         \label{fig:clay_refresh}
     \end{subfigure}
     \begin{subfigure}[b]{0.24\textwidth}
         \centering
         \includegraphics[width=\textwidth]{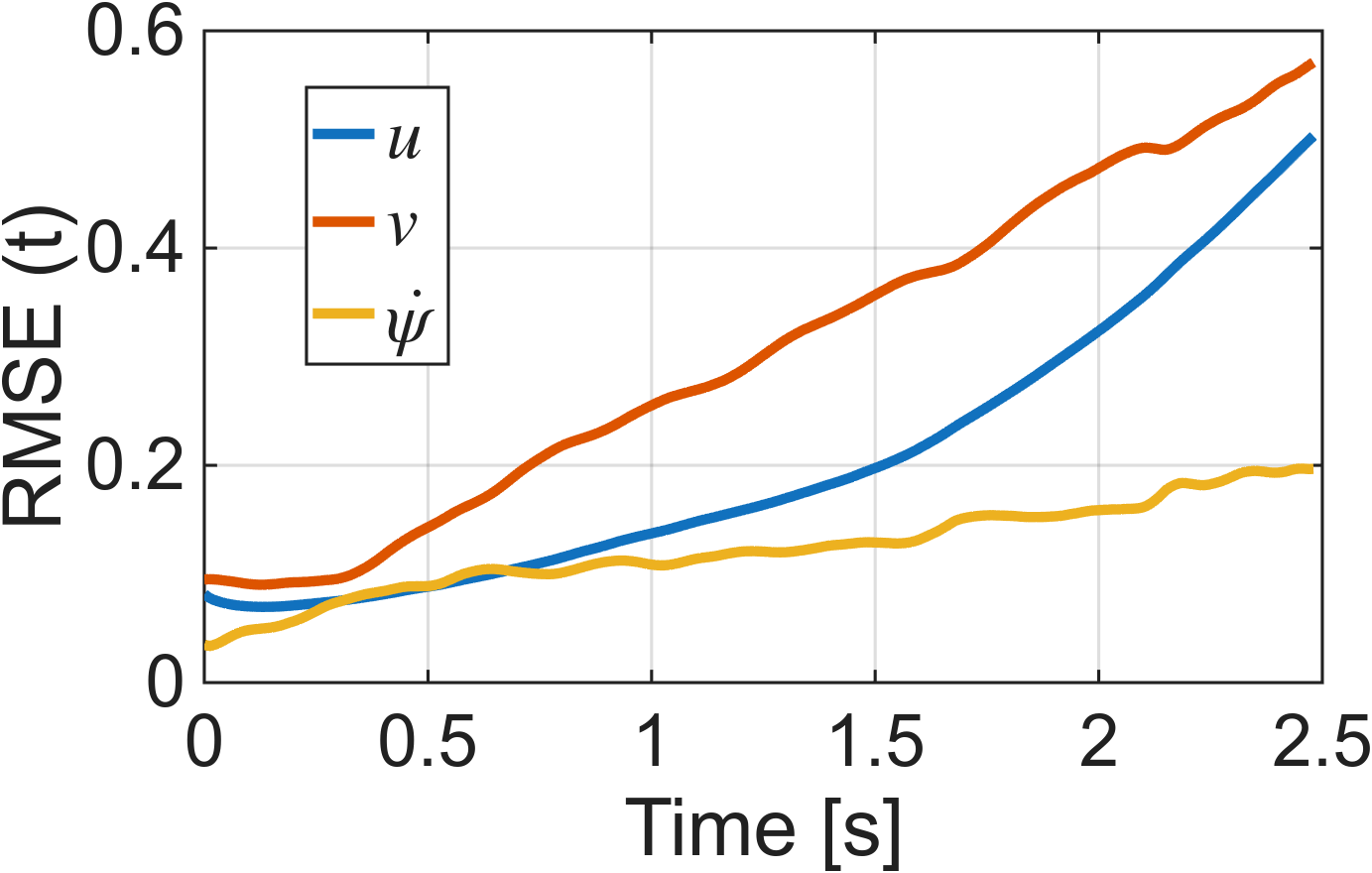}
         \caption{}
         \label{fig:clay_err_time}
     \end{subfigure}
          \begin{subfigure}[b]{0.24\textwidth}
         \centering
         \includegraphics[width=\textwidth]{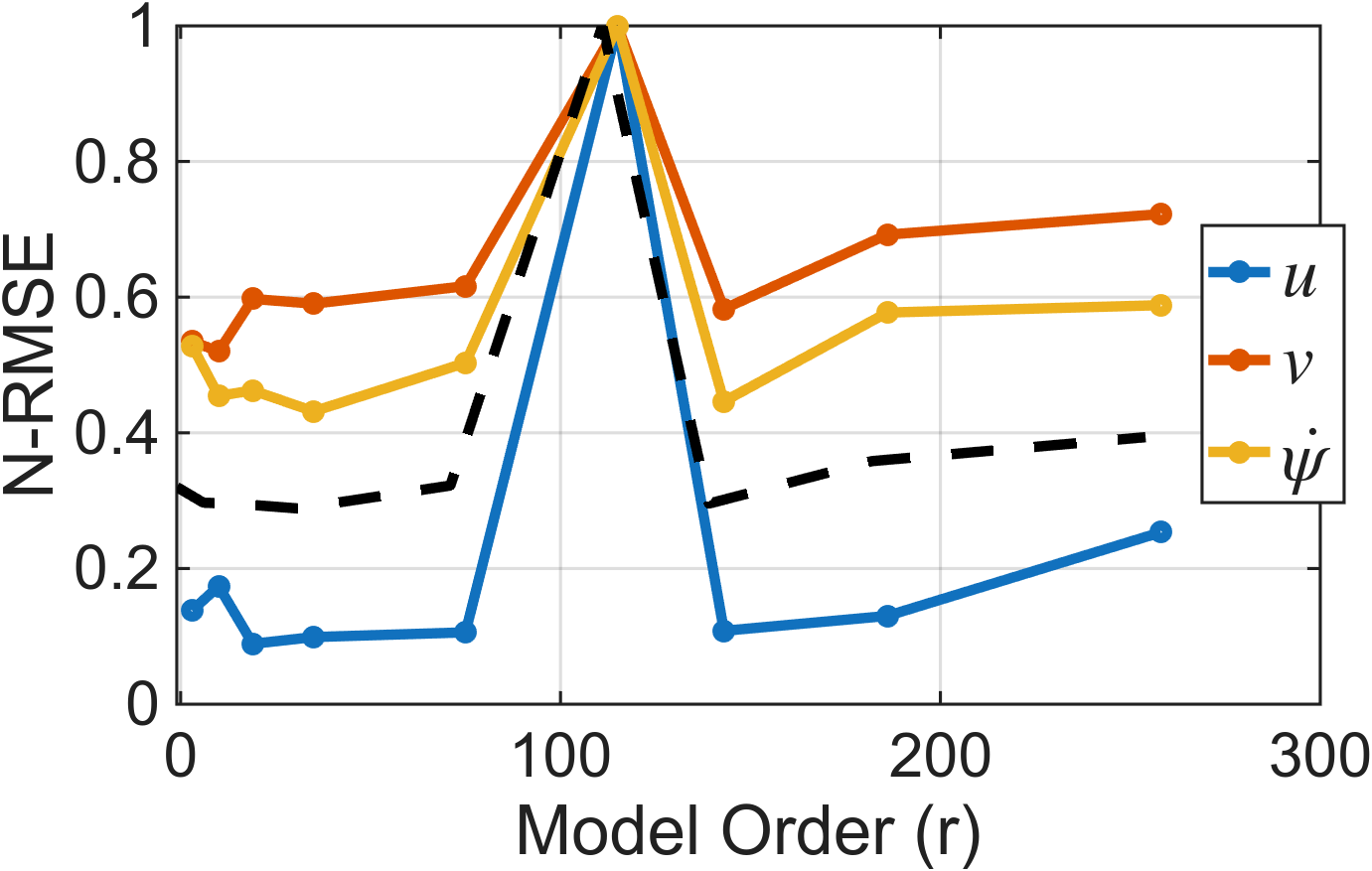}
         \caption{}
         \label{fig:clay_MC}
     \end{subfigure}
          \begin{subfigure}[b]{0.24\textwidth}
         \centering
         \includegraphics[width=\textwidth]{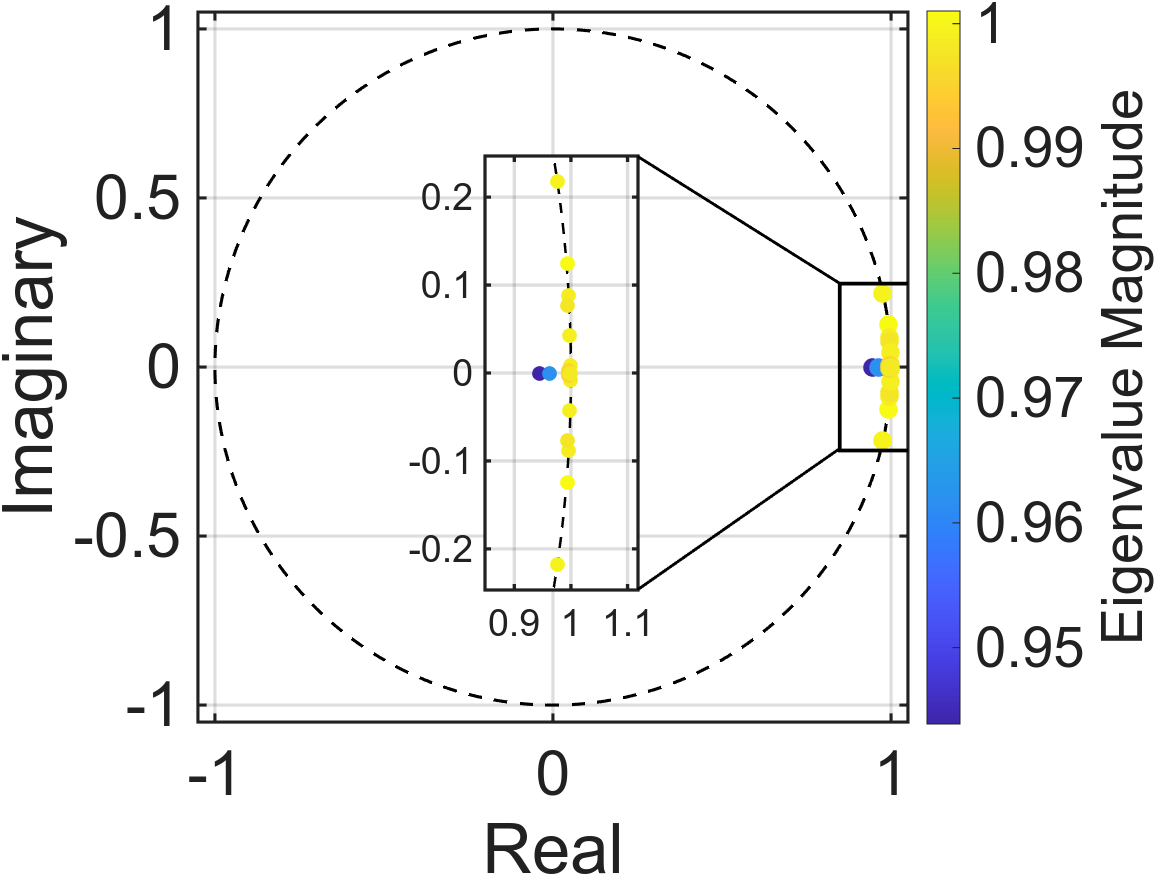}
         \caption{}
         \label{fig:clay_eigen}
     \end{subfigure}
     \caption{Clay soil results for Koopman model selection and prediction: (a) RMSE versus refresh time, (b) open-loop RMSE growth with time, (c) normalized RMSE versus model order $r$ and (d) eigenvalue spectrum of the identified Koopman $A$ matrix.}\label{fig:clay_results}
\end{figure}

\begin{figure}[t]
        \centering
        \includegraphics[width=0.48\textwidth, keepaspectratio]{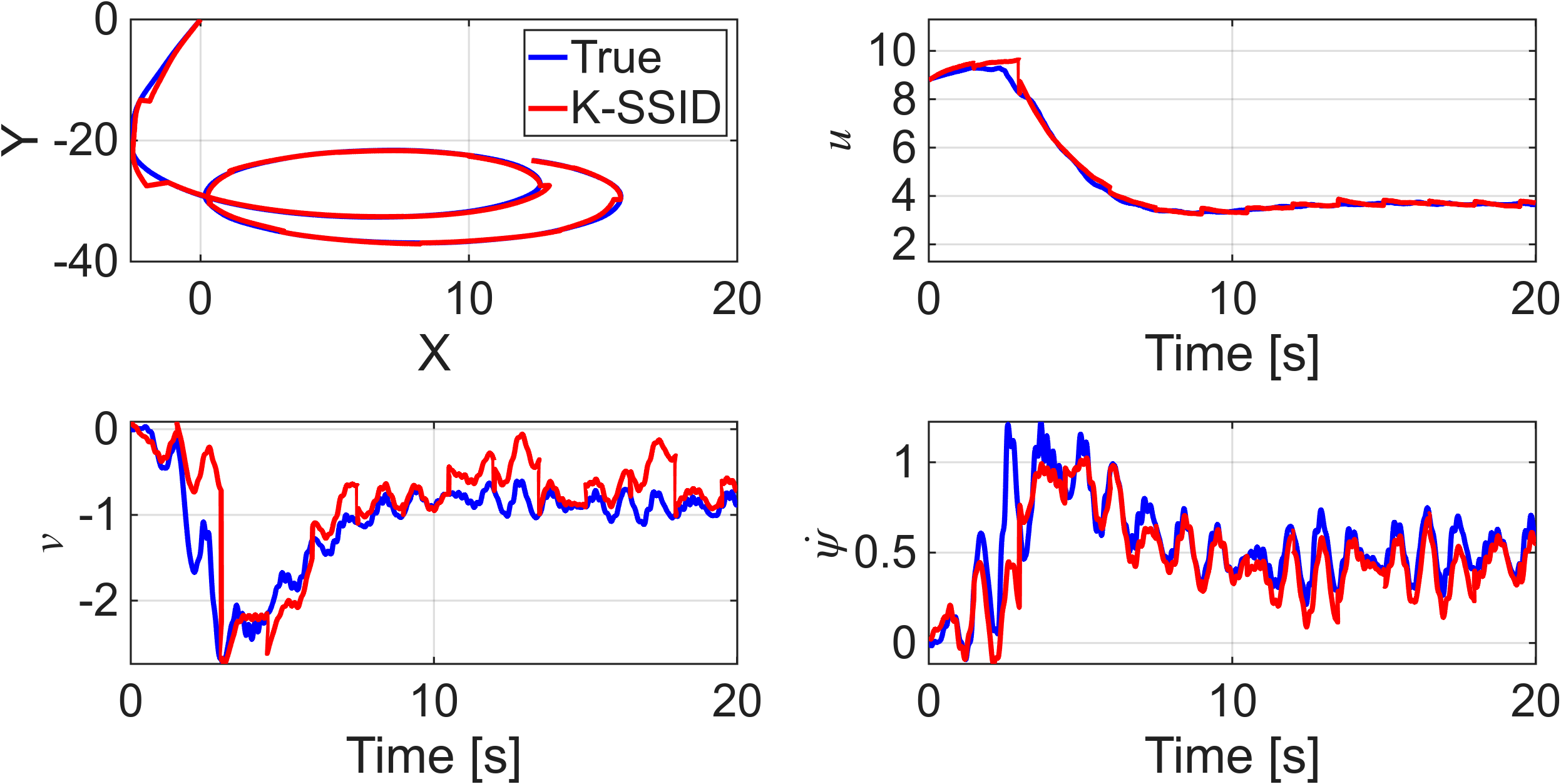}
        \caption{Sample trajectory prediction on clay soil.} 
        \label{fig:clay_traj_pred}
\end{figure}

where $\mathcal{R}$ defines the set of all models computed for that soil with various model order. The normalized errors are shown for sandy loam in Fig.~\ref{fig:SL_MC} and for clay in Fig.~\ref{fig:clay_MC}, where the dotted black line denotes the mean normalized error across all outputs. As $r$ increases, the prediction error initially decreases because additional modes better capture the dynamics. However, beyond a certain order the improvement becomes marginal and the normalized error becomes less consistent across outputs, suggesting diminishing returns as additional low-energy modes are weakly identified and may lead to overfitting. Fig.\ref{fig:SL_refresh} for sandy loam and Fig.~\ref{fig:clay_refresh} for clay soil show the RMSE error over different refresh times from $0.25s$ to $2.5 s$ which gives an insight into the accumulation of errors with respect to the prediction horizon. Among the outputs, $v$ typically degrades the fastest, while $u$ and $\dot\psi$ remain comparatively better predicted over the same horizon. Now, in the Fig.~\ref{fig:SL_err_time} and Fig.~\ref{fig:clay_err_time}, the RMSE shown is at each time step for open-loop predictions over a $2.5s$ segment for sandy loam and clay, respectively, which is calculated as,
\begin{equation}
RMSE_y(t) = \frac{1}{N_t}\sum_{i=1}^{N_t} \Bigg(\sqrt{(\hat{y}_{t} - y_{t})^2} \Bigg),  
\end{equation}
 The results show that the prediction error grows with time when no refresh is applied, with 
$v$ increasing faster than others, consistent with the horizon-dependent trends observed previously. The eigenvalue plot for sandy loam Koopman modeling Fig.~\ref{fig:SL_eigen} and for clay soil in Fig.~\ref{fig:clay_eigen} confirms that the identified linear dynamics are stable $(\lambda_{eig} < 1)$ and clustered around $1$, which in the discrete-time Koopman indicate slow, weakly damped modes, a characteristic of inertia-dominated vehicle dynamics relative to the sampling interval $\Delta t = 0.01$. The model order for the sandy loam Koopman model is $18$ and for the clay model is $19$. Finally, the example trajectory prediction highlight the qualitative prediction capability of the soil-specific Koopman models. For sandyloam, the curved-trajectory example in Fig.~\ref{fig:SL_traj_pred} shows close agreement with the ground truth, capturing both the overall path evolution and the dominant oscillatory content in the lateral and yaw dynamics. Likewise, the clay example in Fig.~\ref{fig:clay_traj_pred} demonstrates stable prediction over a more aggressive spiral maneuver, reproducing the overall path shape and the major lateral–yaw transients with modest mismatch primarily during high-curvature segments where wheel-terrain nonlinearity is strongest. For both the example prediction stated use a refresh time of $2.5 s$ over a $20 s$ trajectory, that is, the predictor is periodically re-initialized every $2.5 s$ using the true outputs.

\subsection{Model Generalization Under Terrain Elevation Perturbations and Maneuver Diversity}
\begin{figure}[t]
        \centering
        \includegraphics[width=0.48\textwidth, keepaspectratio]{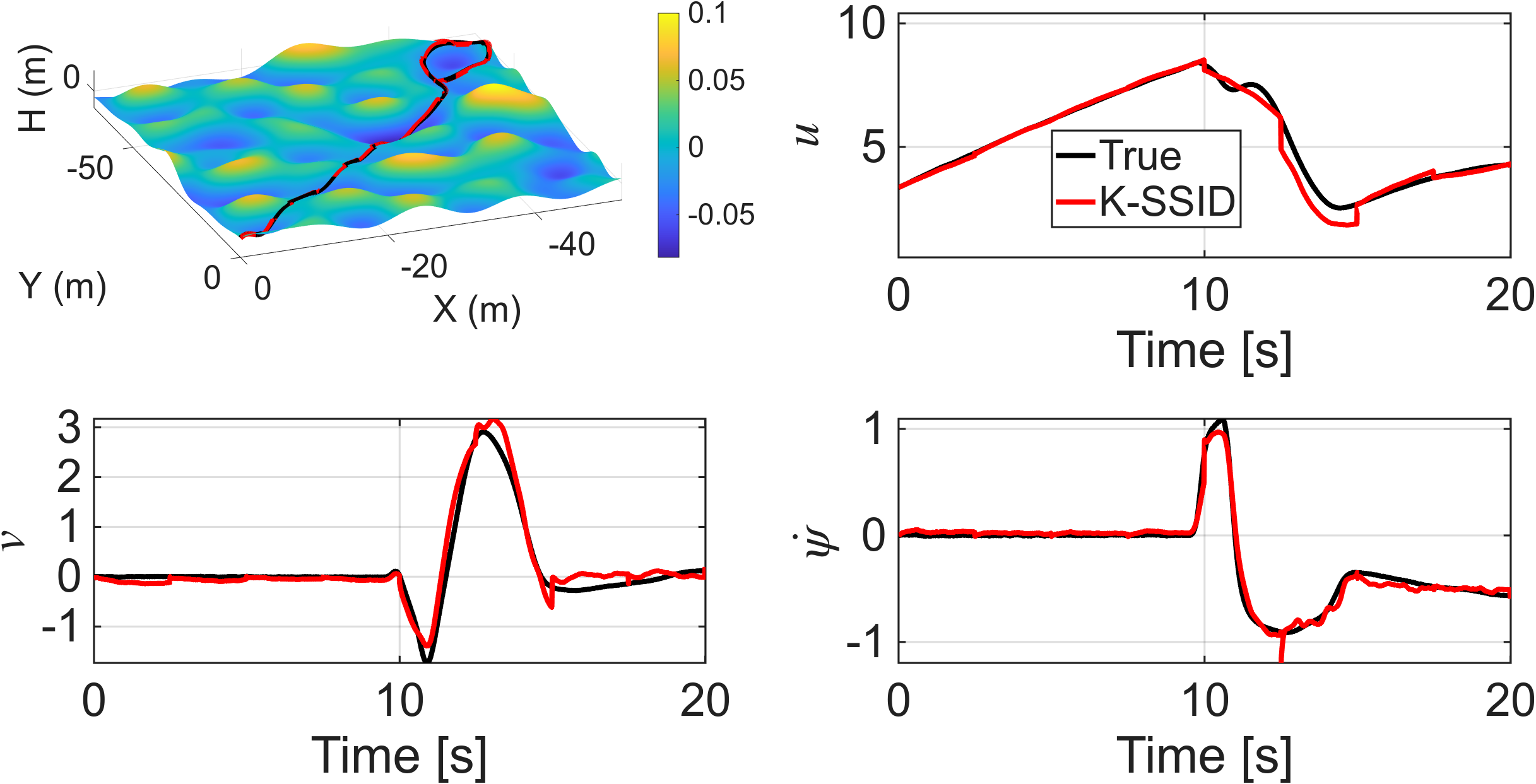}
        \caption{Sample trajectory prediction on sandy loam with terrain elevation of maximum $0.1m$ height.}
        \label{fig:SL_elev_traj_pred}
\end{figure}

\begin{figure}[t]
        \centering
        \includegraphics[width=0.48\textwidth, keepaspectratio]{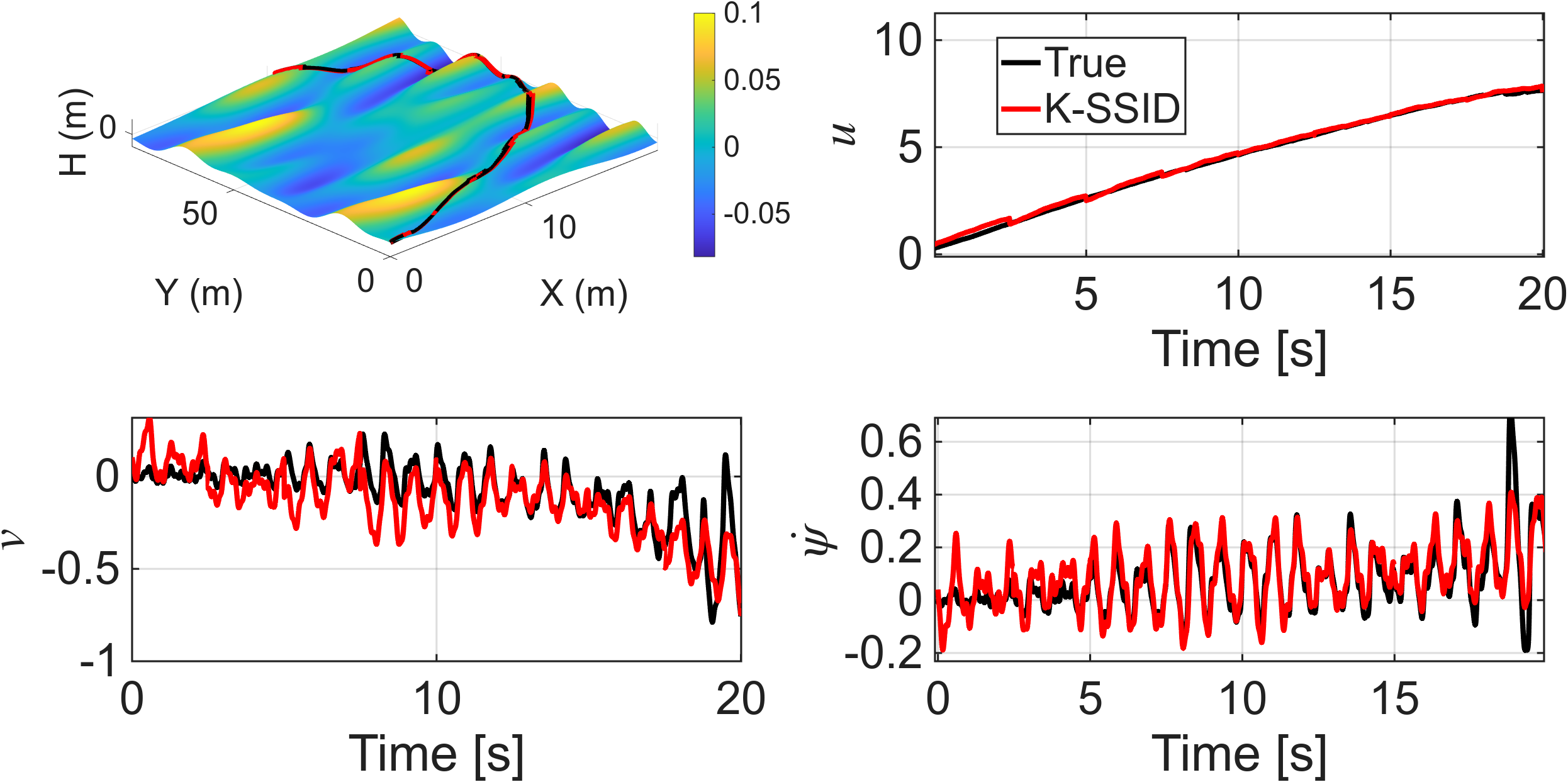}
        \caption{Sample trajectory prediction on clay with terrain elevation of maximum $0.1m$ height.} 
        \label{fig:clay_elev_traj_pred}
\end{figure}

The results in Fig.~\ref{fig:sandyloam_results}–\ref{fig:clay_traj_pred} were obtained on flat terrain. To assess robustness, we evaluated the same learned Koopman predictors, trained only on flat-ground data, on trajectories generated over a mildly uneven terrain with a maximum elevation variation of $0.1m$, without retraining. The sandy loam example in Fig.~\ref{fig:SL_elev_traj_pred} shows a fishhook maneuver on the height map, and the clay example in Fig.~\ref{fig:clay_elev_traj_pred} shows a curved slalom on the height map. In both cases, the predictor remains stable and provides close agreement with the ground truth for the planar motion, lateral and longitudinal speed and yaw rate indicating that the learned Koopman dynamics are robust to these small terrain-height variations. To quantify the effect of terrain variation, we compute the percentage increase in RMSE for the elevated-terrain test dataset relative to the RMSE of flat-terrain test dataset  with a $2.5 s$ refresh time for Koopman model prediction is reported in Table~\ref{tab:elev_rmse}. The $\%RMSE_y$ is given as 
\begin{equation}
    \%RMSE_y = 100*\frac{RMSE_y(elev) - RMSE_y(no~elev)}{RMSE_y(no~elev)}
\end{equation}

\begin{figure}[h]
     \begin{subfigure}[b]{0.24\textwidth}
         \centering
         \includegraphics[width=\textwidth]{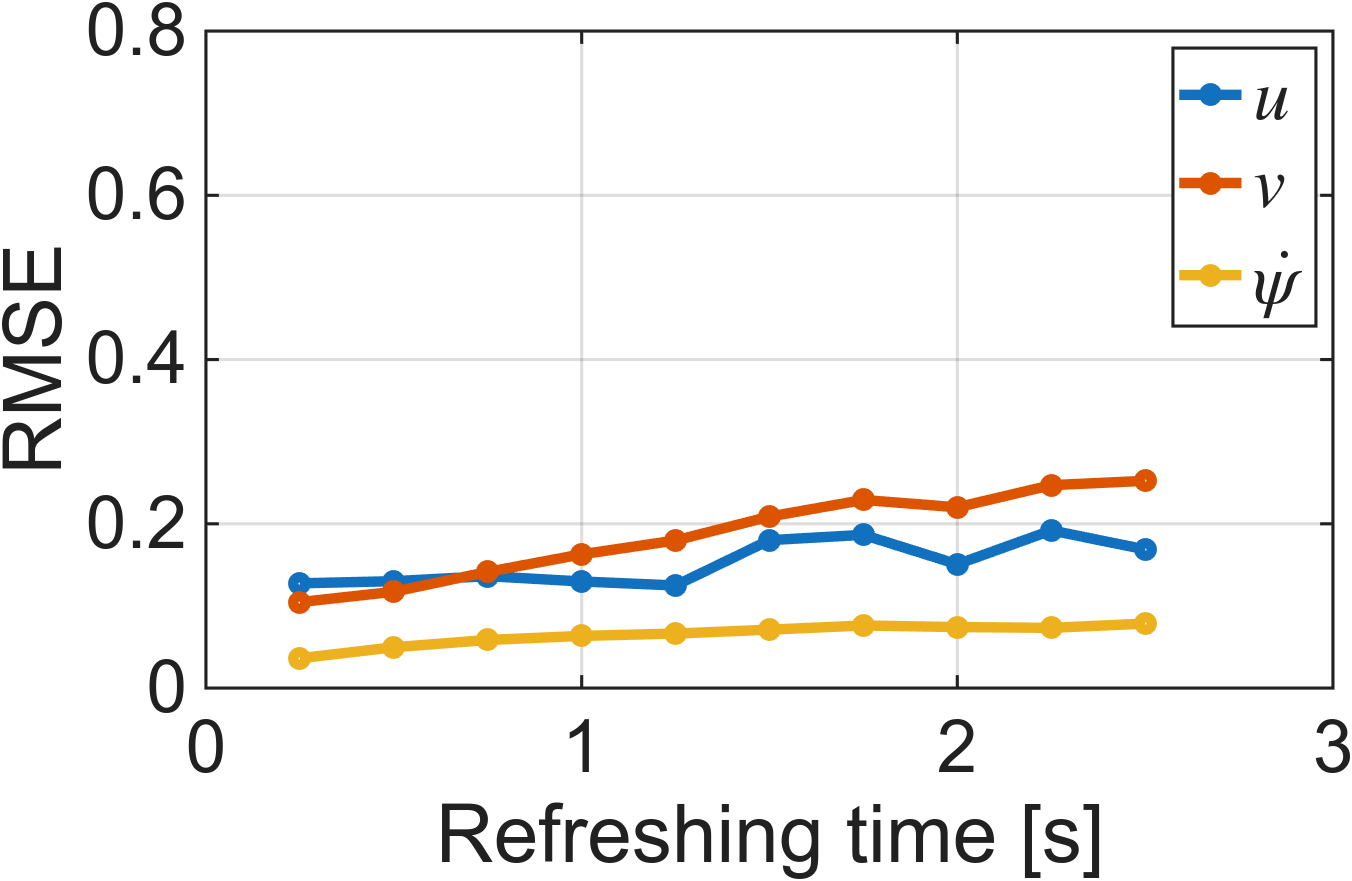}
         \caption{Sandy loam soil}
         \label{fig:sl_elev_refresh}
     \end{subfigure}
     \begin{subfigure}[b]{0.24\textwidth}
         \centering
         \includegraphics[width=\textwidth]{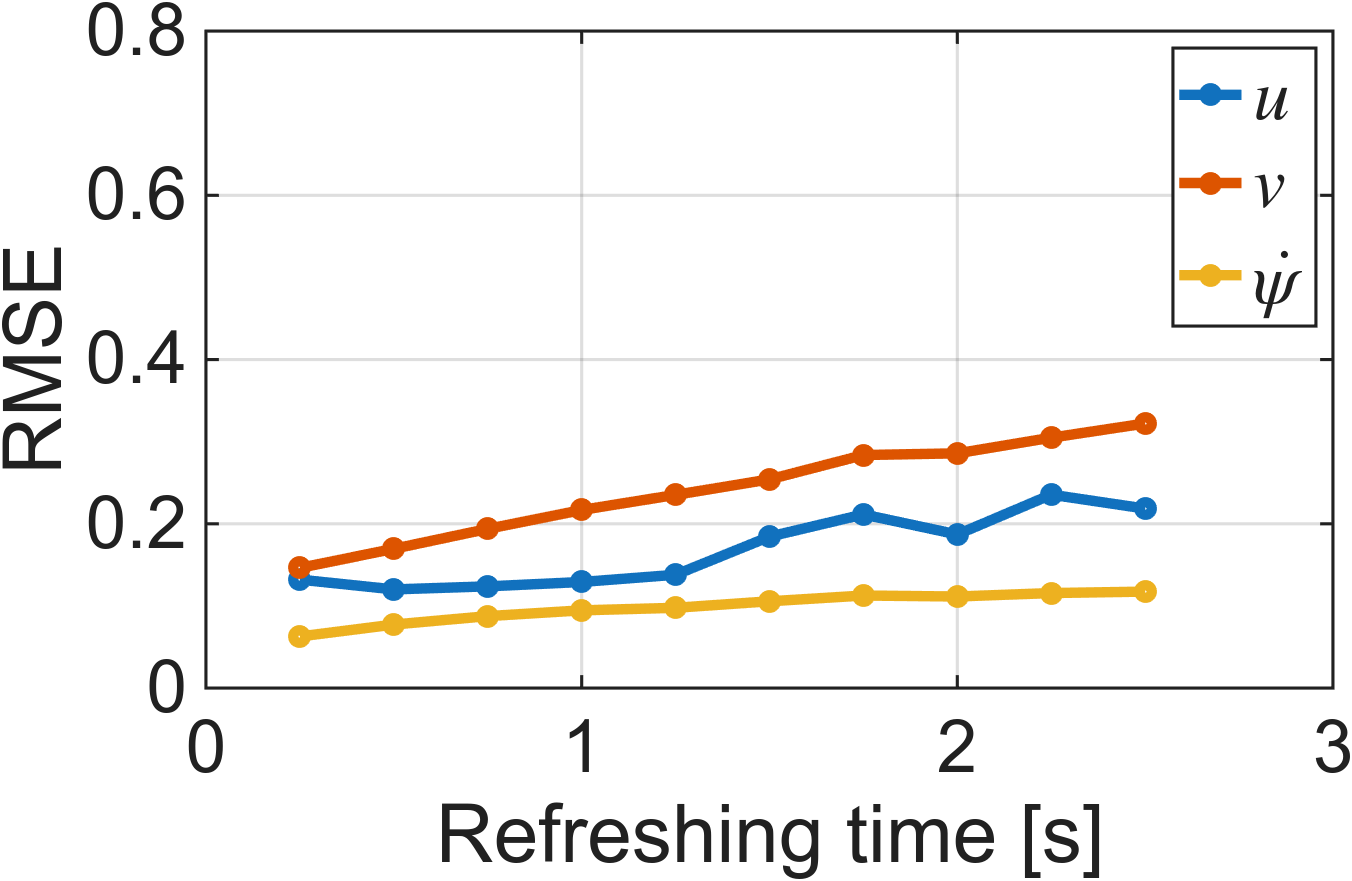}
         \caption{Clay soil}
         \label{fig:clay_elev_refresh}
     \end{subfigure}
     \caption{(a) RMSE versus refresh time for sandy loam soil with terrain elevation, (b) RMSE versus refresh time for clay soil with terrain elevation.}\label{fig:refresh_elev}
\end{figure}

\begin{table}[h]
\caption{RMSE percentage increased for dataset with elevation with refresh rate $1.25 s$}\label{tab:elev_rmse}
\setlength{\tabcolsep}{3pt} 
\centering{%
\begin{tabular}{c c c c}
\toprule
Soil & $\% RMSE_u$ & $\% RMSE_v$ & $\% RMSE_{\dot{\psi}}$ \\
\midrule
Sandy Loam &  $14.6048$  &  $9.4656$  &  $3.2462$ \\
Clay &  $15.0323$   &  $2.2215$ &  $9.8556$   \\
\bottomrule
\end{tabular}
}%
\end{table}

Table~\ref{tab:rmse_maneuvers} reports the prediction RMSE for the sandyloam and clay Koopman models across five maneuver classes (straight, circle, multisine, slalom, and fishhook). For each maneuver, the RMSE values were computed over 50 test trajectories for the reported states $[u,v,\dot{\psi} ]$. The results indicate that the training dataset provides good coverage of each maneuver class, enabling a consistent comparison across maneuvers and soils. Overall, the sandy loam model exhibits consistently lower errors and smaller variation across maneuvers, with $u$ remaining near $0.26$, $v$ near $0.29$, and $\dot{\psi}$ near $0.064$. In contrast, the clay model shows higher RMSE across all maneuvers, with $u$ around $0.30$, $v $ around $0.37$, and 
and $\dot{\psi}$ around $0.156$. Across both soils, the maneuver-to-maneuver differences are modest compared to the overall soil-to-soil gap, indicating that terrain type has a stronger effect on prediction accuracy than the specific maneuver class for these test conditions.
\begin{table}[h]
\caption{RMSE on different maneuvers}\label{tab:rmse_maneuvers}
\setlength{\tabcolsep}{3pt} 
\centering{%
\begin{tabular}{l l| l l c l l}
\toprule
Soil & & Straight & Circle & Multisine & Slalom & Fishhook  \\
\midrule
Sandy&$u$ & 0.260  &  0.260 &   0.256  &  0.257  &  0.259 \\
loam&$v$ & 0.292  &  0.285 &   0.287  &  0.287  &  0.290 \\
&$\dot{\psi}$ & 0.064  &  0.063 &   0.064  &  0.064  &  0.064 \\
\midrule
Clay&$u$ & 0.296  &  0.311  &  0.315  &  0.310  &  0.303 \\
&$v$ &  0.390 &  0.373  &  0.376  &  0.382  &  0.377 \\
&$\dot{\psi}$ & 0.155  &  0.152  &  0.155  &  0.158  &  0.157 \\
\bottomrule
\end{tabular}
}\label{tab:rmse_maneuvers}
\end{table}

Across these tests, the clay model exhibits higher error than the sandy loam model, primarily because the clay dataset contains larger lateral-velocity excursions and a wider spread of slip ratio, which increases the variability of the wheel-terrain forces.
In contrast, the sandy loam dataset shows a comparatively tighter operating range in slip and lateral motion, leading to more consistent wheel-terrain relationships and lower prediction error. 

\section{Koopman-Based Model Predictive Control}\label{sec:kmpc}

This section formulates a Koopman-based MPC scheme for time-varying trajectory tracking using the learned linear predictor. The Koopman model provides a lifted linear evolution of the vehicle body-frame dynamics, while a kinematic update is used to propagate the global position states.

\subsection{Prediction model}\label{subsec:kmpc_model}
Let $z_k \in \mathbb{R}^{r}$ denote the Koopman lifted state at time step $k$, and let the control input be
$u_k = [\delta_k,\;\tau_k]^\top$, where $\delta$ is steering and $\tau$ is drive torque. The identified discrete-time Koopman predictor is
\begin{equation}
z_{k+1} = A z_k + B u_k,
\label{eq:kmpc_koopman}
\end{equation}
where $A\in\mathbb{R}^{r\times r}$, $B\in\mathbb{R}^{r\times 2}$. The body-frame outputs are recovered via
\begin{equation}
y_k = C z_k, 
\qquad
y_k = [u_k^{b},\; v_k^{b},\; \dot{\psi}_k]^\top,
\label{eq:kmpc_output}
\end{equation}
where $C\in\mathbb{R}^{3\times r}$. Here, $u_k^{b}$ and $v_k^{b}$ are the longitudinal and lateral velocities in the body frame, and $\dot{\psi}_k$ is yaw rate.

To obtain the global planar motion, we augment the predictor with kinematic states
$x_k = [X_k,\;Y_k,\;\psi_k]^\top$ and integrate them using the predicted body-frame velocities:
\begin{align}
X_{k+1} &= X_k + \big(u_k^{b}\cos\psi_k - v_k^{b}\sin\psi_k\big)\Delta t, \label{eq:kmpc_kin1}\\
Y_{k+1} &= Y_k + \big(u_k^{b}\sin\psi_k + v_k^{b}\cos\psi_k\big)\Delta t, \label{eq:kmpc_kin2}\\
\psi_{k+1} &= \psi_k + \dot{\psi}_k \Delta t, \label{eq:kmpc_kin3}
\end{align}
where $\Delta t$ is the MPC discretization time step. The full tracked output is then
\begin{equation}
\tilde{y}_k = [X_k,\;Y_k,\;\psi_k,\;u_k^{b},\;v_k^{b},\;r_k]^\top.
\label{eq:kmpc_fulloutput}
\end{equation}

\subsection{MPC optimization problem}\label{subsec:kmpc_opt}
At each MPC update, we solve a finite-horizon constrained optimization over a prediction horizon $N_p$. Let $\tilde{y}_{\mathrm{ref},k}$ denote the reference trajectory over the horizon. The MPC problem is
\begin{align}
    \min_{\{u_k\}} \;\; & 
    \sum_{k=0}^{N_p-1} \Bigg( \big(\tilde{y}_k-\tilde{y}_{\mathrm{ref},k}\big)^\top Q \big(\tilde{y}_k-\tilde{y}_{\mathrm{ref},k}\big) \;+\; u_k^\top R\,u_k \;+\; \label{eq:kmpc_cost} \\
     &~~~~~~~~~~~~~~~\Delta u_k^\top R_{\Delta u}\,\Delta u_k,
     \Bigg) \nonumber \\
    \text{s.t.}\;\; &
    z_{k+1} = A z_k + B u_k, \qquad k=0,\dots,N_p-1, \label{eq:kmpc_dyn1}\\
    &
    y_k = C z_k + c_c, \label{eq:kmpc_dyn2}\\
    &
    x_{k+1} \text{ satisfies \eqref{eq:kmpc_kin1}--\eqref{eq:kmpc_kin3}},  \label{eq:kmpc_dyn3}\\
    &
    u_{\min} \le u_k \le u_{\max},  \label{eq:kmpc_bounds}
\end{align}
where $Q\succeq 0$ penalizes tracking error in $\tilde{y}_k$, $R\succeq 0$ penalizes control effort, and $R_{\Delta u}\succ 0$ penalizes input rate. In our implementation, the bounds are $\delta \in [-0.35,\,0.35]$ and $\tau \in [0,\,130]$.

\subsection{State initialization and receding-horizon implementation}\label{subsec:kmpc_impl}
Since the Koopman lifted state $z_k$ is not directly measured, the initial lifted state $z_0$ is obtained from a Gaussian-process lifting map trained offline. Specifically, using the measured subset of outputs $y = [u, v, \dot{\psi}]$, we predict each component of $z_0$ using the learned GPR models from Eq.\eqref{eq:GP}:
\begin{equation}
z_0 = \psi(y), 
\end{equation}
where $\psi(\cdot)$ denotes the GP-based lifting map. At each MPC update, the nonlinear vehicle model provides the current state from which the measured output is fed to the GP to lift it to $z_0$, and the optimization problem \eqref{eq:kmpc_cost}--\eqref{eq:kmpc_bounds} is solved using IPOPT through CasADi. The controller then applies the first portion $(N_c)$ of the optimal control sequence in a receding-horizon manner, and the process repeats with updated measurements.
\begin{figure}[ht]
     \begin{subfigure}[b]{0.48\textwidth}
         \centering
         \includegraphics[width=\textwidth]{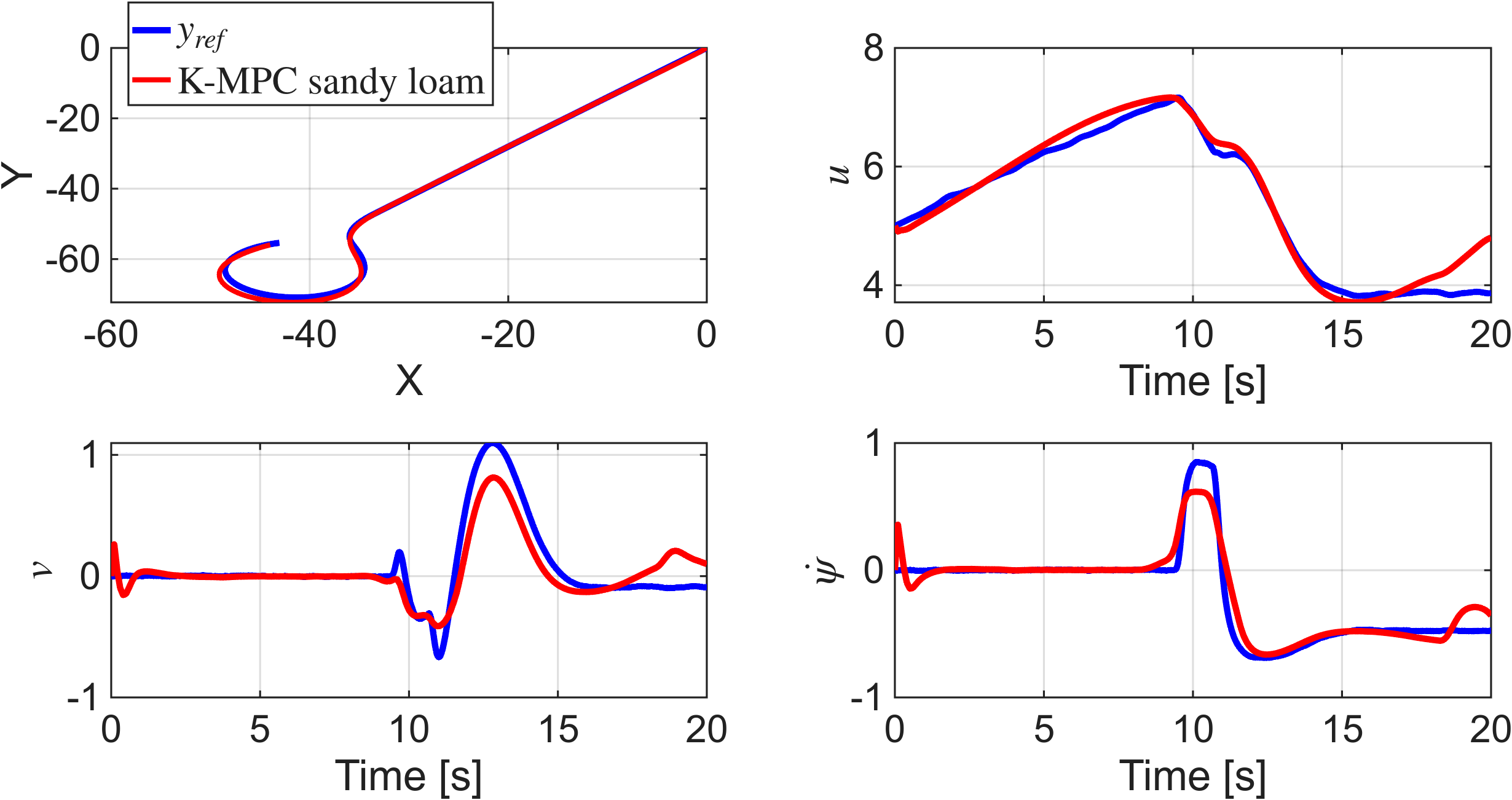}
         \label{fig:mpc_sl_1}
     \end{subfigure}\\
     \begin{subfigure}[b]{0.48\textwidth}
         \centering
         \includegraphics[width=\textwidth]{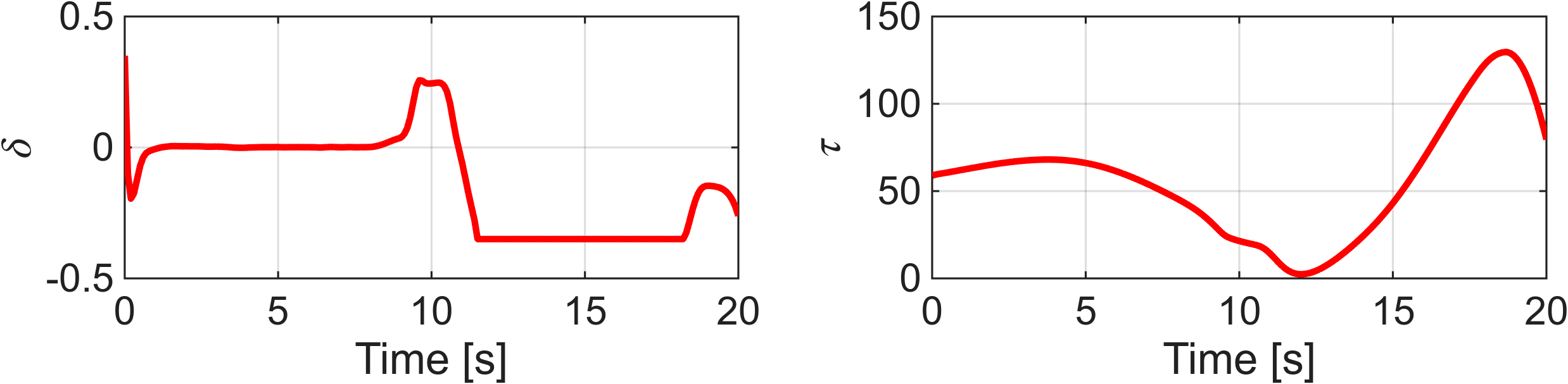}
         \label{fig:mpc_sl_2}
     \end{subfigure}
     \caption{Koopman-based MPC tracking on sandyloam. Top: reference trajectory $y_{\mathrm{ref}}$ and executed trajectory under K-MPC in the $(X,Y)$ plane, and the corresponding state tracking for $u$, $v$, and $\dot{\psi}$. Bottom: optimized control inputs steering $\delta$ and drive torque $\tau$ applied by the MPC.}
\label{fig:sandy_kmpc}
\label{fig:mpc_sl_2}
\end{figure}

In the experiments, the MPC discretization is set to $\Delta t = 0.1\,\mathrm{s}$. A prediction horizon of $N_p=20$, corresponding to a $2\,\mathrm{s}$ look-ahead window. The receding-horizon controller is updated every control horizon, for example, $N_c=5$ applies the first $0.5\,\mathrm{s}$ segment of the optimized input sequence before resolving the MPC with updated measurements. This choice leverages the observation from the prediction analysis that short refresh times of $0.5\,\mathrm{s}$ maintain bounded error while retaining sufficient horizon length for planning and tracking. The quadratic objective penalized tracking error using  $Q=diag([15,15,15,1,15,15])
$, lightly penalized control magnitude using 
$R=diag([10^{-2},10^{-6}])$, and enforced smooth inputs using the input rate penalty 
$R_{\Delta u}=diag([100,1])$. Fig.~\ref{fig:sandy_kmpc} presents the closed-loop trajectory-tracking result for a fishhook maneuver on sandy loam using the proposed Koopman-based MPC. The fishhook is a highly maneuverable and dynamically challenging test because it introduces rapid changes in curvature and strong lateral-yaw transients. The planar motion in the $(X,Y)$ plane shows that the controller follows the reference path closely. The time histories indicate that the controller reduces the longitudinal speed $u$ prior to the high-curvature portion of the maneuver and then executes the turn while maintaining stable tracking of the dominant lateral and yaw dynamics. The transient behavior in the lateral velocity $v$ and the yaw rate $\dot{\psi}$ is captured with only small deviations. The corresponding control inputs, steering $\delta$ and drive torque $\tau$, adapt to regulate both heading and speed while satisfying the imposed actuator limits. Overall, these results demonstrate that the learned Koopman predictor is sufficiently accurate for receding-horizon control and enables stable tracking performance on sandy loam. Moreover, the optimization ran in real time, with a mean solver time of $0.085~\text{s}$ per MPC iteration, remaining below the sampling period of $\Delta t = 0.1~\text{s}$.


\section{Need for Terrain-Specific Koopman Operators}\label{sec:6}

\begin{figure}[h]
        \centering
        \includegraphics[width=0.34\textwidth, keepaspectratio]{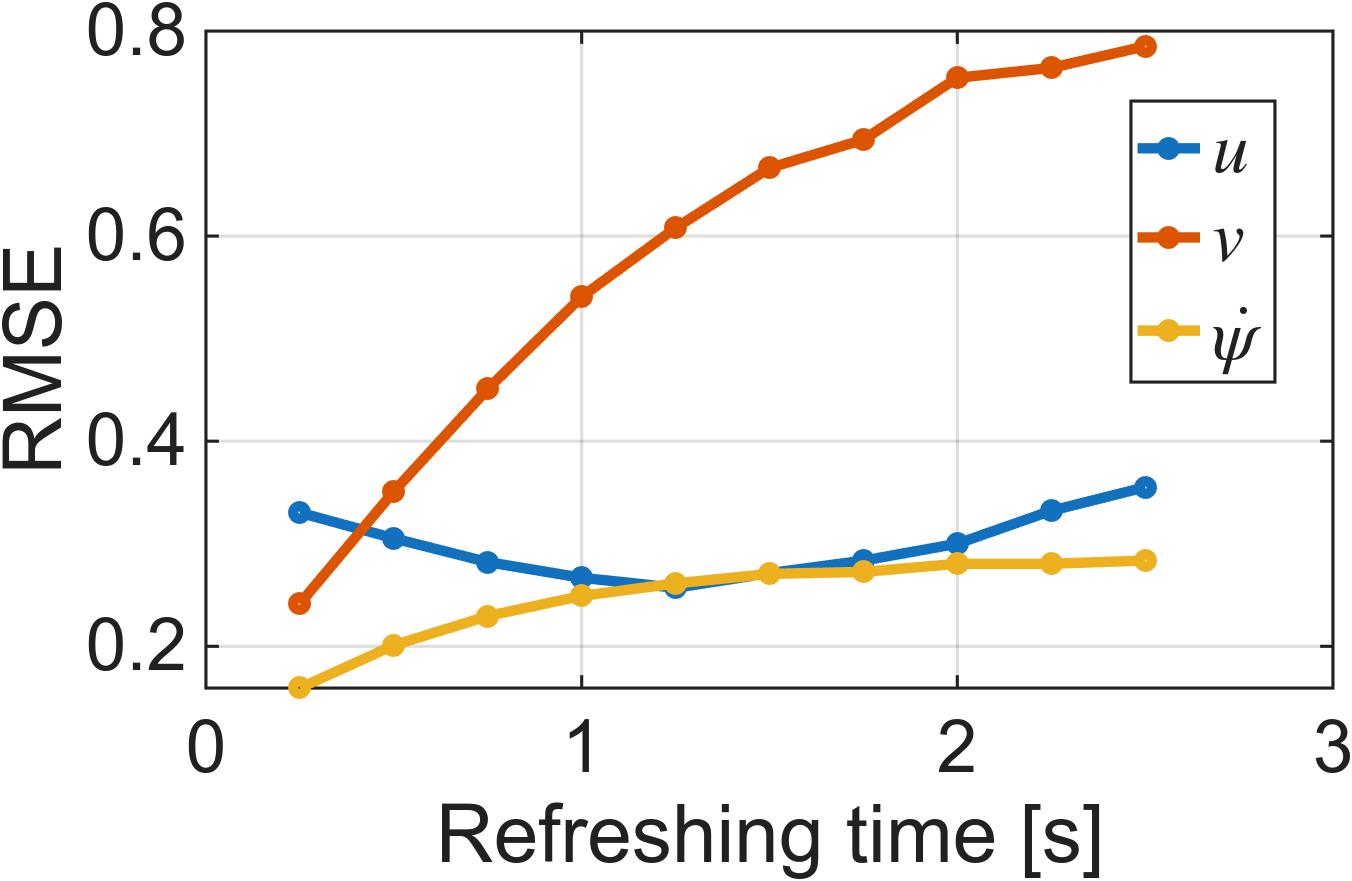}
        \caption{RMSE versus refresh time for sandy loam Koopman operator on clay soil.}
        \label{fig:sl_on_clay_refresh}
\end{figure}
On deformable terrain, the closed loop response is governed not only by vehicle kinematics but also by soil dependent wheel terrain interaction, which changes the mapping from steering and drive commands to lateral velocity and yaw dynamics. To quantify this dependence, we evaluated cross terrain prediction on a clay test dataset using two Koopman operators, one identified from sandy loam data and one identified from clay data. The clay trained operator maintained low prediction error across all refresh times, see Fig.\ref{fig:clay_refresh}, whereas the sandy loam operator showed a systematic mismatch that increased with refresh time, see Fig.\ref{fig:sl_on_clay_refresh}. Using the sandy loam operator on clay increased RMSE by $90$ percent for $u$, $137$ percent for $v$, and $149$ percent for $\dot{\psi}$ on average across all refresh time, relative to the clay trained operator. 
This gap is not due to controller tuning. It arises because the vehicle dynamics change with soil type. The largest errors occur in $v$ and $\dot{\psi}$, which are most sensitive to soil shear and slip, indicating that the cross terrain mismatch is driven by wheel–terrain interaction. 

\begin{figure}[h]
     \begin{subfigure}[b]{0.49\textwidth}
         \centering
         \includegraphics[width=\textwidth]{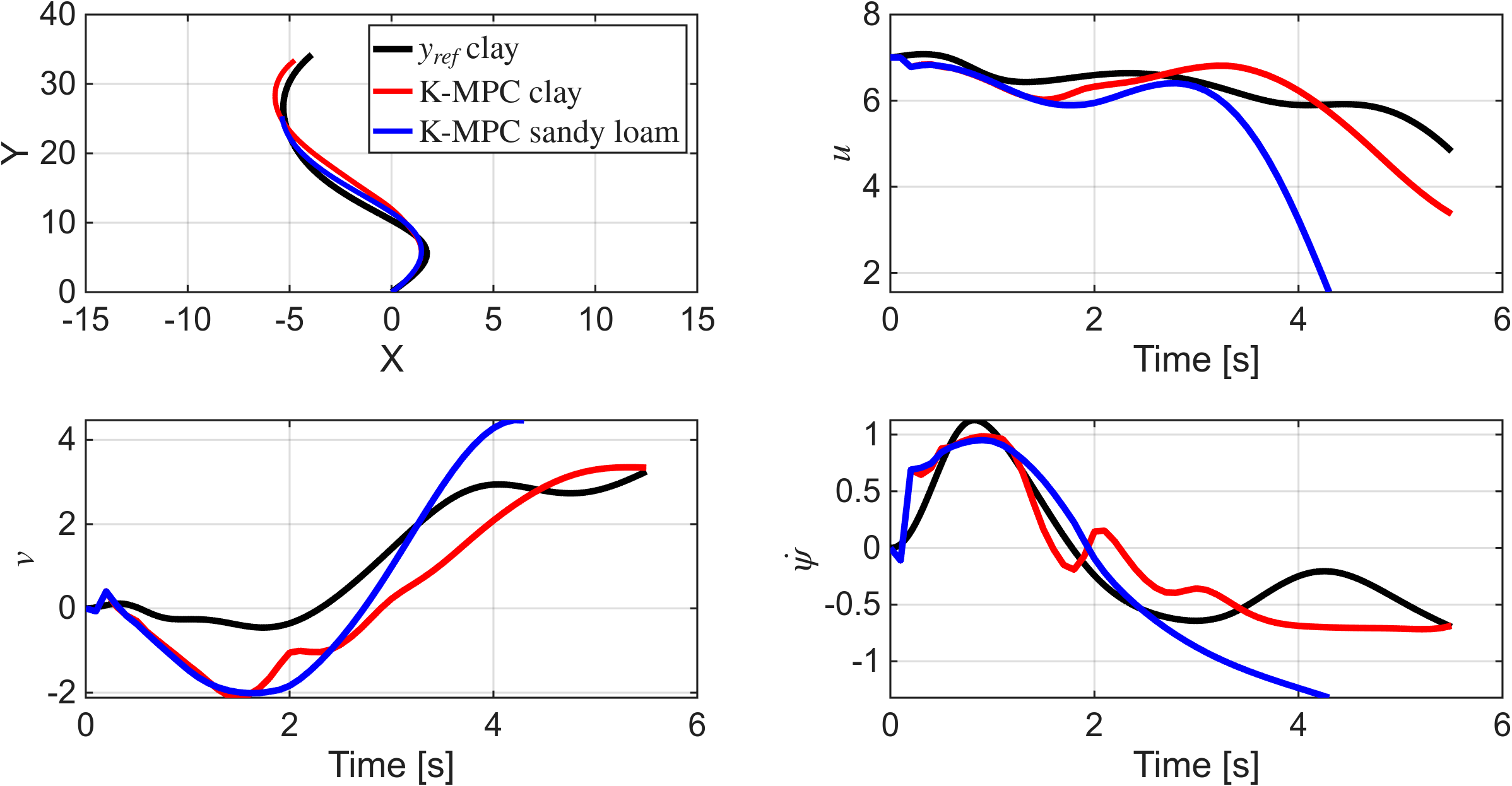}
         \label{fig:mpc_dual_1}
     \end{subfigure} \\
     \begin{subfigure}[b]{0.49\textwidth}
         \centering
         \includegraphics[width=\textwidth]{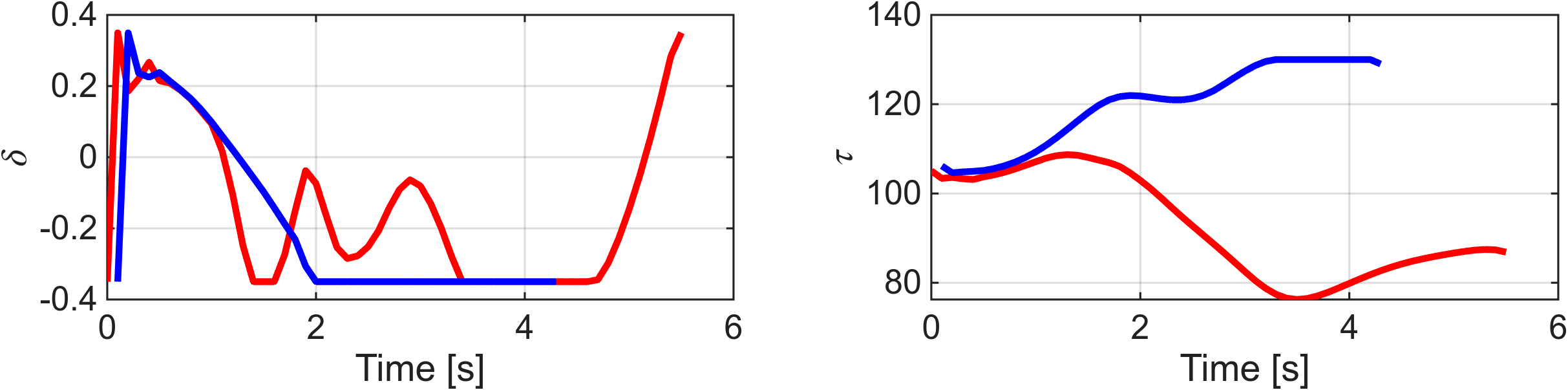}
         \label{fig:mpc_dual_2}
     \end{subfigure} 
     \caption{Terrain based Koopman operator MPC tracking on clay soil. Top: reference trajectory $y_{\mathrm{ref}}$ and executed trajectory under sandy loam K-MPC (blue) and clay K-MPC (red). Bottom: optimized control inputs steering $\delta$ and drive torque $\tau$ applied by the MPC.}
\label{fig:dual_kmpc}
\end{figure}

Building on the prediction error results above, we next compared closed loop MPC tracking on the same clay terrain while changing only the learned Koopman predictor used inside the optimizer. The MPC was implemented with a prediction horizon of 
$Np=20$ steps and sampling time 
$\Delta t=0.1s$, which corresponds to a $2 s$ look ahead window, and a short control horizon of 
$Nc=5$ steps. The quadratic MPC objective used the same weighting matrices as in the previous example, penalizing tracking error, control effort, and control-rate variations to enforce smooth inputs. With the clay-trained Koopman operator, the MPC tracks the reference path closely and the state trajectories remain consistent with the desired evolution. In particular, the lateral velocity $v$ and yaw-rate $\dot{\psi}$ follow the reference trends, and the steering input $\delta$ makes small corrective adjustments, including mild oscillations, to maintain the tracking direction without persistent saturation. This behavior yields a substantially lower mean objective value of $1042$ per iteration. 
\begin{figure}[h]
    \centering\includegraphics[width=0.34\textwidth, keepaspectratio]{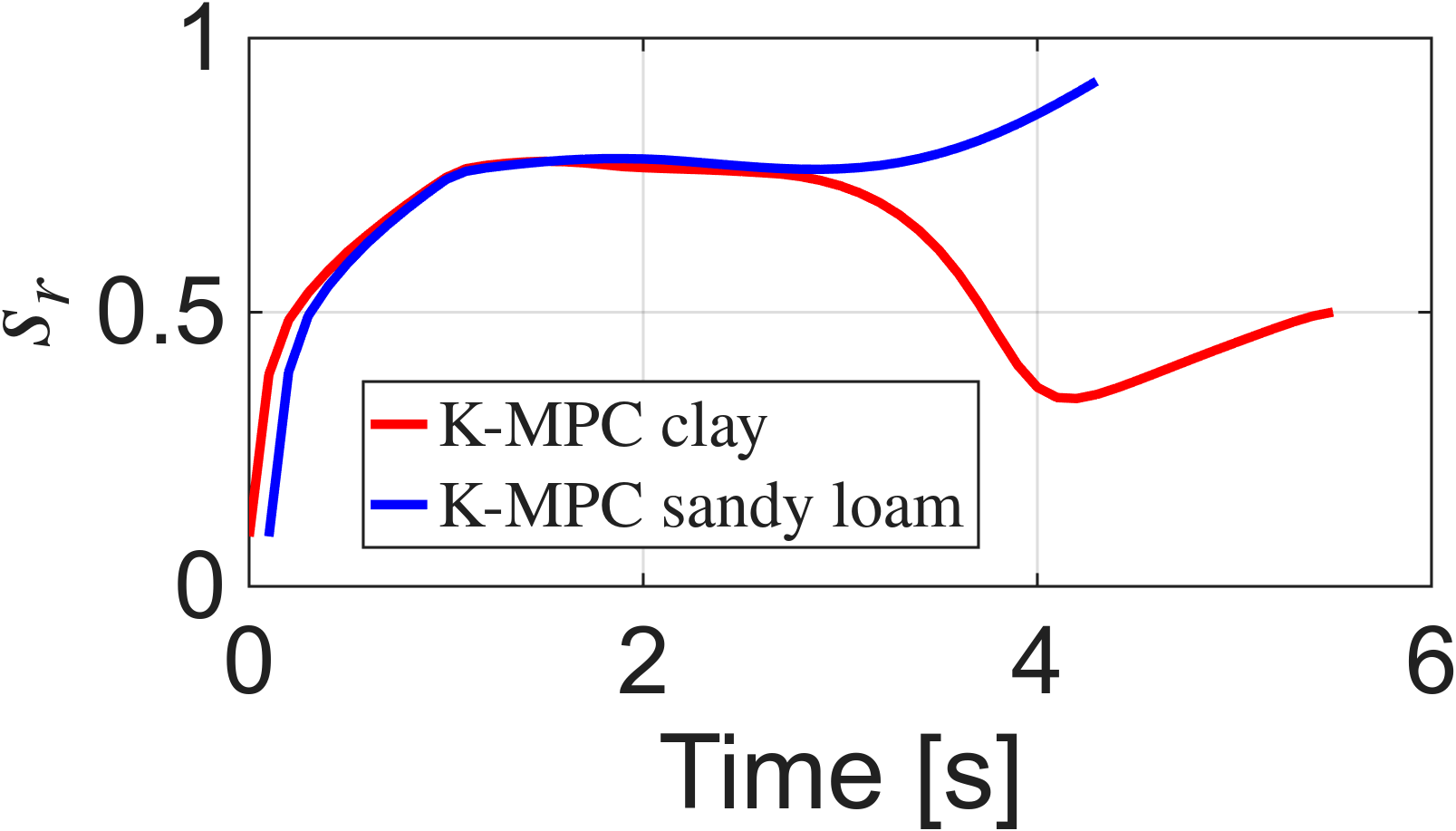}
        \caption{Slip ratio for sandy loam K-MPC and Clay K-MPC in rear wheel.}
        \label{fig:MPC3_dual_slip}
\end{figure}
In contrast, when the sandy loam trained Koopman operator is used on the clay terrain, the closed-loop response degrades earlier and the vehicle drifts away from the reference, accumulating larger tracking error over the maneuver. The time histories show larger discrepancies in $v$ and $\dot{\psi}$, and the slip-related response departs from the clay-trained case see Fig.\ref{fig:MPC3_dual_slip}, indicating that the optimizer is planning with an inaccurate wheel-terrain interaction model. As the mismatch grows, the MPC drives the steering input $\delta$ to the steering bounds for extended periods, but this constraint-limited actuation cannot recover the trajectory because the internal predictor misrepresents the clay lateral dynamics. In our implementation, the simulation is stopped when the steering remains saturated and the vehicle can no longer follow the reference trajectory. The degraded tracking is captured quantitatively by the increased mean objective value of $2929$ per iteration, nearly three times larger than the clay-trained case, despite identical horizons, weights, and constraints. 

Overall, these results show that cross-terrain prediction errors propagate directly into MPC decisions, leading to higher control effort, increased slip, higher cost, and earlier trajectory drift or failure when the Koopman operator does not match the terrain. These observations motivate terrain-specific Koopman operators and, more generally, online operator selection or adaptation mechanisms, since an operator identified on one soil does not reliably generalize when wheel-soil effects dominate the motion response.

\section{Conclusions}\label{sec:7}
This paper presents a data-driven Koopman framework for approximating the nonlinear dynamics of off-road vehicles resulting from strong wheel–terrain interactions. The goal is to develop a simplified linear model that can predict key dynamic states while still capturing terrain-dependent effects that are challenging to model with conventional low-order models. Using high-fidelity simulations with Bekker-based wheel–terrain forces, we generated datasets with varied maneuvers for sandy loam and clay soil and identified separate soil-specific Koopman models using a recursive subspace identification (K-SSID) \cite{LBT_dsc_2023, LT_mecc_2024} approach. 
Prediction results showed stable short-horizon accuracy, with model order selected using a singular-value sweep and RMSE validation. Using periodic refresh of the predictor state to limit error growth over longer rollouts, the flat-terrain-trained models also maintained good performance on mildly uneven terrain with up to $0.1m$ elevation variation. Maneuver-wise RMSE was consistent across straight, circle, multisine, slalom, and fishhook tests within each soil type. Finally, embedding the learned Koopman predictor in a constrained MPC achieved stable tracking of an aggressive maneuver while respecting steering and torque limits, demonstrating real-time receding-horizon control. More broadly, the Koopman lifted-linear predictor offers a computationally efficient alternative to high-fidelity terramechanics simulation for short-horizon prediction and control on soft soil while retaining the dominant dynamics. 

Future work can extend the approach to larger terrain height variations and additional soil conditions, and also study ensembles of Koopman models that are adaptive to soil conditions and can be selected online or blended to handle mixed soil properties by weighting multiple operators using Grassmannian distance or prediction error. More interestingly for future work, the approach in this paper can be generalized to a combined physics informed data driven calculation of Koopman operator. The Koopman operator for a vehicle can be calculated using simulations, either using Bekker or Wong-Reece models or a high fidelity simulation environment like Project Chrono. These simulations are sensitive to terrain parameters which can be uncertain. In-situ measurements and estimation of these parameters is possible as a vehicle traverses the uncertain terrain. The prior computed Koopman operator can be updated based on time-series data collected from a vehicle in an efficient manner.

\section*{Acknowledgment} 
This work was supported by Clemson University’s Virtual Prototyping of Autonomy Enabled Ground Systems (VIPR-GS), under Cooperative Agreement W56HZV-21-2-0001 with the US Army DEVCOM Ground Vehicle Systems Center (GVSC).

\section*{Publication Distribution Statement}
DISTRIBUTION STATEMENT A. "Approved for public release; distribution is unlimited. OPSEC \# 10375"














\bibliographystyle{asmejour}   

\bibliography{offroad_Koopman} 



\end{document}